\documentclass[12pt]{iopart}
\usepackage{iopams}
\expandafter\let\csname equation*\endcsname\relax
\expandafter\let\csname endequation*\endcsname\relax
\usepackage{graphicx}
\usepackage{times}

\newcommand{\qqq}{\end{document}}
\DeclareMathAlphabet{\mathcal}{OMS}{cmsy}{m}{n}
\usepackage{xcolor}

\usepackage{bbding}
\usepackage{mathrsfs}
\usepackage{braket}
\usepackage[protrusion=true,expansion=true]{microtype}
\usepackage{cite}
\usepackage{dsfont}
\usepackage{centernot}
\usepackage{color,amsthm,amsmath,amsxtra,amsfonts,dsfont,graphicx,bm}
\usepackage[colorlinks=true,linkcolor=blue, citecolor=blue, urlcolor=blue, bookmarks]{hyperref}
\usepackage{graphicx}
\usepackage{tikz}
\usepackage{braket}
\usepackage{multirow}
\usepackage{comment}
\usepackage{makecell}

\newcommand{\de}{\partial}
\newcommand{\E}{\mathbb{E} }
\newcommand{\Var}{{\rm Var} }
\newcommand{\be}{\begin{equation}}
	\newcommand{\ee}{\end{equation}}
\newcommand{\bea}{\begin{eqnarray}}
	\newcommand{\eea}{\end{eqnarray}} 
\newcommand{\bse}{\begin{subequations}}
	\newcommand{\ese}{\end{subequations}}
\newcommand{\id}{\bm{1}}
\theoremstyle{plain}

\usepackage{pifont}

\begin{document}
	\date{\today}
	
	\newcommand{\bbra}[1]{\<\< #1 \right|\right.}
	\newcommand{\kket}[1]{\left.\left| #1 \>\>}
	\newcommand{\bbrakket}[1]{\< \Braket{#1} \>}
	\newcommand{\pll}{\parallel}
	\newcommand{\nn}{\nonumber}
	\newcommand{\transp}{\text{transp.}}
	\newcommand{\nor}{z_{J,H}}
	
	\newcommand{\hL}{\hat{L}}
	\newcommand{\hR}{\hat{R}}
	\newcommand{\hQ}{\hat{Q}}
	
	
\title{Domain-wall melting in all-to-all QSSEP from random-matrix theory
}

\author{Denis Bernard~$^1$, \ Lorenzo Piroli~$^{2,\dagger}$, \ Stefano Scopa~$^{1,*}$}

\address{\footnotesize$^1$~Laboratoire de Physique de l’\'Ecole Normale Superieure, CNRS,
ENS \& Universit\'e PSL, Sorbonne Universit\'e, Universit\'e Paris Cit\'e, 75005 Paris, France.\\[4pt]
$^2$~Dipartimento di Fisica e Astronomia, Universit\`a di Bologna and INFN, Sezione di Bologna, via Irnerio 46, 40126 Bologna, Italy.\\[4pt]
$^\dagger$~\href{mailto:lorenzo.piroli@unibo.it}{lorenzo.piroli@unibo.it};\; 
$^*$~\href{mailto:stefano.scopa@phys.ens.fr}{stefano.scopa@phys.ens.fr}
}

\begin{abstract}
We study the melting of a domain wall in the quantum simple exclusion process with all-to-all hoppings (a.k.a. the charged SYK$_2$ model). We show that the real-time dynamics of physical quantities of interest can be obtained exploiting spectral results in random matrix theory. We first show that the eigenvalues of the correlation matrix corresponding to the initially charged subsystem evolve according to a Jacobi process, which is defined in terms of a closed system of stochastic differential equations. In turn, this observation allows us to obtain the real-time dynamics of all the eigenvalue moments. We present two physical applications. First, we study the dynamics of the averaged von Neumann entanglement entropy, arriving at a fully explicit expression in the thermodynamic limit. Second, we compute analytically the full-counting statistics of the charge. Our formula allows us to perform a thorough comparison with the full-counting statistics of the classical simple exclusion process. Notably, we show that, in the thermodynamic limit, the quantum and classical full-counting statistics coincide, with no finite-time corrections. 
\end{abstract}

\maketitle

	{\small\tableofcontents}
	\newpage
\markboth{}{}
\section{Introduction}

 Stochastic processes and random-matrix theory (RMT) are very versatile tools in quantum many-body physics, allowing us to model a number of different physical situations. Besides providing a useful description of open-system dynamics~\cite{breuer2002theory}, where the unknown interactions with the environment introduce effective randomness, stochastic processes can also be used to study the typical behavior of isolated many-body systems. In this context, random variables model generic values of the interactions, allowing one to explore large regions in parameter space and remove any fine tuning. 
 
 Perhaps counterintuitively, modeling the typical dynamics of a many-body system by stochastic processes may introduce simplifications in its theoretical description, even leading to analytic results. In the past decade, this approach has been beautifully exemplified within the study of random unitary circuits (RUC)~\cite{nahum2017quantum,potter_entanglement_2022,fisher2023random}. These models proved to be ideal settings to investigate quantities that are hard to compute away from exactly solvable models, allowing us to obtain analytic results on, e.g., the entanglement growth~\cite{nahum2017quantum,rakovszky2019sub,emergent2019zhou,gullans2019entanglement,bertini2019entanglement,znidarivc2020entanglement,huang2020dynamics,entanglement2020zhou}, the operator spreading~\cite{nahum2018operator,vonKeyserlingk2018operator,chan2018Solution,sunderhauf2018localization,diffusive2018rakovszky,khemani2018operator,hunter2018operator,friedman2019spectral}, and scrambling of quantum information~\cite{hosur2016chaos,scrambling2020bertini,piroli2020random}.

RUC were originally defined as discrete dynamics, but parallel studies considered analogous models defined in terms of stochastic continuous-time Hamiltonian evolution~\cite{bauer2017stochastic,onorati2017mixing,knap2018entanglement,rowlands2018noisy,zhou2019operator,sunderhauf2019quantum,bernard2020entanglement}. In this work, we focus in particular on one model that has played an important role in the past decade, namely the quantum symmetric simple exclusion process (QSSEP)~\cite{bauer2017stochastic,bauer2019equilibrium,barraquand2025introduction}. This model can be defined on different graphs, and it describes fermions hopping with random amplitudes between sites that are connected by an edge of the graph. As such, it can be viewed as a quantization of the SSEP, a prototypical model for classical transport~\cite{kipnis1989hydrodynamics,spohn2012large,derrida1993exact,derrida2007non,bodineau2004current,mallick2015exclusion}. 

The QSSEP introduces simplifications compared to RUC models, as its dynamics is described in terms of an Hamiltonian generator that is quadratic in the fermionic modes, allowing us to employ analytic techniques from the theory of fermionic Gaussian states~\cite{bravyi2004lagrangian,surace2022tagliacozzo}. At the same time, despite its apparent simplicity, the model and its generalizations~\cite{bernard2019open,jin2020stochastic,alba2025nu,bernard2025large} display a rich phenomenology, making them ideal toy models to build a quantitative understanding of quantum fluctuations, and a solid basis towards the development of a quantum macroscopic fluctuation theory (MFT) for diffusive transport~\cite{bertini2001fluctuations,bertini2002macroscopic,bertini2015macroscopic}. 

In the past few years, several analytic results have been derived to describe the late-time properties of the QSSEP, both in the absence of an external bath~\cite{bauer2019equilibrium,bernard2021solution,bernard2021entanglement,alba2025nu,russotto2026dynamics,russotto2026inhomogeneous} and in the presence of transport-inducing boundary driving~\cite{bernard2019open,hruza2023coherent,bernard2023exact,bauer2024bernoulli,bernard2025large,albert2026universal}. On the contrary, it is highly non-trivial to characterize the QSSEP dynamics. Indeed, while the averaged time evolution of simple observables is mapped to the classical SSEP~\cite{bauer2017stochastic}, the dynamics of quantities that are non-linear in the system density matrix is determined by coherent quantum fluctuations. Examples of such quantities include the bipartite entanglement entropy and the particle full-counting statistics, describing the particle fluctuations within a given spatial region. 

A standard approach to study the moments of coherent quantum fluctuations in the QSSEP is based on the mapping onto a non-unitary evolution in a suitable replica space~\cite{nahum2017quantum,bernard2022dynamics}. This approach allowed, for instance, for the exact computation of the subsystem purity~\cite{swann2025spacetime}, which involves second moments of the density matrix, but becomes more complicated for the analysis of quantities involving higher moments~\cite{bernard2022dynamics}. This difficulty represents a bottle-neck to arrive at a complete analytic understanding of the quantum-fluctuation dynamics.

In this work, we consider a non-equilibrium protocol involving the \emph{melting} of an initially localized domain wall, focusing on the fully connected QSSEP, also known as the charged SYK$_2$ (after Sachdev, Ye, and Kitaev~\cite{sachdev1993gapless,maldacena2016remarks}). Recent results for the integrability of a family  of clean SYK models can be found in Ref.~\cite{Fukai2026}. Although this is a special setting,  domain-wall melting is a standard protocol where to gain an intuition on various phenomena, including transport processes and entanglement growth~\cite{Antal1999,Antal2008,Karevski2002,Platini2007,Hunyadi2004,Collura2018,Gruber2019,Eisler2018,Scopa2021,Ares2022, capizzi2023domain, Scopa2023,russotto2026inhomogeneous}. As our main result, we develop an analytic approach describing the melting dynamics, based on recent spectral results from RMT. We stress that the dynamics of the SYK$_2$ model can also be tackled via the standard replica approach, to obtain predictions for the dynamics of quantities like the system purity~\cite{tiutiakina2025field}. Our approach differs from previous studies in that it gives us direct access to all moments of the coherent quantum fluctuations. 

Specifically, we first show that the eigenvalues of the correlation matrix corresponding to the initially charged subsystem evolve according to a Jacobi process, which is defined in terms of a closed system of stochastic differential equations (SDE). In turn, this observation allows us to obtain the real-time dynamics of all the eigenvalue moments. We present two physical applications. First, we study the dynamics of the averaged von Neumann entanglement entropy, arriving at a fully explicit expression in the thermodynamic limit. Second, we compute analytically the full-counting statistics of the charge. The interest of our results is two-fold. On the one hand, our work presents a new approach to stochastic dynamics, featuring a particularly nice and direct application of RMT constructions. On the other hand, our analytic formulas allow for a rare exact comparison between particle fluctuations in the quantum and classical SSEP dynamics.\\

\emph{Outline}~---~The rest of this work is organized as follows. In Sec.~\ref{sec:the_model}, we introduce the fully-connected QSSEP and the domain-wall quench protocol, and recall how entanglement and particle fluctuations can be computed from the spectrum of the correlation matrix. In Sec.~\ref{sec:jacobi_process}, we show that the non-zero eigenvalues of the reduced correlation matrix evolve according to a Jacobi process, and we exploit known results from RMT to characterize their dynamics in the thermodynamic limit.  In Sec.~\ref{sec:phys-app} we apply these results to compute explicitly the time evolution of the von Neumann entanglement entropy and of the charge full-counting statistics. In Sec.~\ref{sec:classical-ssep} we compare the latter with the corresponding classical dynamics of the all-to-all SSEP, showing that the stationary fluctuations coincide in the thermodynamic limit. In Sec.~\ref{sec:finite-time-comparison} we extend this comparison to finite times, deriving the evolution equations for the generating functions and quantifying finite-size corrections. Finally, Sec.~\ref{sec:outlook} contains our conclusions and perspectives, while technical details are collected in the appendices.

\section{Setup}
\label{sec:the_model}
\subsection{The model and the quench protocol}

We consider a model of $L$ fermionic modes and study the all-to-all QSSEP (or charged Brownian SYK$_2$ model), defined by the Hamiltonian generator
\be\label{eq:dH}
d\hat{H}(t)=\frac{1}{\sqrt{L}} \sum_{1\leq i< k \leq L}\left( \hat c^\dagger_j \hat c_k \ dW^{jk}(t) + \hat c^\dagger_k \hat c_j \ d\overline{W}^{jk}(t)\right)\,.
\ee
Here $\{\hat c_j, \hat c^\dagger_k\}=\delta_{jk}$ are standard spinless fermionic operators, while  $dW^{jk}(t)$ and $d\overline{W}^{jk}(t)$ form a pair of complex conjugated Brownian motions satisfying $dW^{jk}(t)d\overline{W}^{mn}(t)=\delta_{jm}\delta_{kn} dt$, where we used the standard notation in It\={o} calculus~\cite{oksendal2003stochastic}. We set the lattice spacing $a=1$ and we rescaled the overall amplitude of the dynamics by a factor $1/\sqrt{L}$ such that the thermodynamic limit yields well-defined quantities. Note that we may rewrite the generator in matrix form as
\be
d\hat{H}(t)=\sum_{i,j=1}^{L-1} \hat c^\dagger_{j} \ \big(dW(t)\big)_{\!\!jk} \ \hat c_k
\ee
where
\be\label{eq:dw_mat}
\big(dW(t)\big)_{\!\!jk}=\begin{cases}
\tfrac{1}{\sqrt{L}} dW^{jk}(t), \quad k\geq j;\\[12pt]
\tfrac{1}{\sqrt{L}} d\overline{W}^{jk}(t), \quad k< j,
\end{cases}
\ee
while
\be\label{eq:unitarity_w}
dW(t) dW(t) =\bm{1}_{L\times L} dt.
\ee

The dynamics of the system is governed by the SDE
\begin{align}
d\hat\rho(t)&= -i[d\hat{H}(t),\hat\rho(t)] -\frac12 [d\hat{H}(t),[d\hat{H}(t), \hat\rho(t)]]=  -i[d\hat{H}(t),\hat\rho(t)] + {\cal L}(\hat\rho) dt
\end{align}
with (deterministic) Lindblad part
\be
{\cal L}(\bullet):= -\frac12 [d\hat{H}(t),[d\hat{H}(t), \bullet]]=\frac{1}{L} \sum_{1\leq i\leq k \leq L} \left(\hat\ell_{jk} \bullet \hat\ell_{jk}^\dagger + \hat\ell^\dagger_{jk} \bullet \hat\ell_{jk}-\frac12\{\hat\ell_{jk}  \hat\ell_{jk}^\dagger + \hat\ell_{jk}^\dagger  \hat\ell_{jk}, \bullet\}\right)
\ee
and jump operators $\ell_{jk}:= \hat c^\dagger_j \hat c_k$ from site $j\to k$ (and symmetrically from $k\to j$).

As mentioned, we focus on a particular nonequilibrium protocol. Namely, we consider preparing the system in a pure state with $M$ particles
\be\label{eq:DW_state}
|\Psi(0)\rangle= \hat c^\dagger_{1} \dots \hat c^\dagger_{M} \  |0\rangle\,,
\ee
and let the system evolve unitarily according to the all-to-all QSSEP. The nonequilibrium protocol can be viewed as a melting of the initial domain wall~\eqref{eq:DW_state}. In the following, we will denote by $\mathcal{U}(t)$ the (stochastic) unitary operator mapping the initial state $\ket{\Psi(0)}$ to the evolved state $\ket{\Psi(t)}$, namely
\begin{equation}\label{eq:unitary_ev}
    \ket{\Psi(t)}=\mathcal{U}(t)\ket{\Psi(0)}\,.
\end{equation}

We will be interested in the evolution of the subsystem $A_\ell=\{1,\dots,\ell\}$, containing the initially charged modes, and we will focus on two quantities. First, we will study the evolution of the bipartite entanglement (R\'enyi) entropy
\be
S_\ell^{(q)}(t):=\frac{\log \ {\rm tr}\left(\hat \rho_\ell(t)^q\right)}{1-q},
\ee
where
\be
\hat\rho_\ell (t)={\rm tr}_{L\setminus A_\ell} |\Psi(t)\rangle\langle \Psi(t)|.
\ee
Note that the Von Neumann entanglement entropy
\be
S_{\ell}(t)=-\ {\rm tr}(\hat \rho_\ell(t)\log \hat\rho_\ell(t) )
\ee
is obtained via the limit $S_\ell(t):=\lim_{q\to 1} S_\ell^{(q)}(t)$. 
Second, we will study particle fluctuation statistics via the moment generating function
\be
Z_\ell(\phi,t):=\langle e^{ i \phi \hat N_\ell} \rangle, 
\ee
where
\begin{equation}
\hat N_\ell=\sum_{j\in A_\ell} \hat c^\dagger_j \hat c_j\,.    
\end{equation}
\begin{figure}[t]
\centering
\includegraphics[width=.65\textwidth]{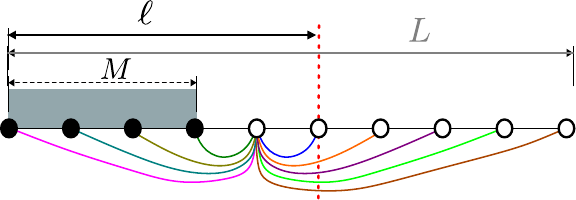}
\caption{Illustration of the setup. We consider a one-dimensional quantum system of size $L$, where spinless fermions can hop between any pair of sites $j\to k$, with amplitudes given by different realization of the complex Brownian motion. The hopping amplitudes are reset at each time step. The system is initialized in a domain-wall configuration, with the leftmost $M$ sites occupied and the rest of the chain empty. During the evolution, we probe fluctuations in a subregion $A_\ell$ of size $\ell$. }\label{fig:setup}
\end{figure}
An illustration of the setup is shown in Fig.~\ref{fig:setup}.
\subsection{Entanglement and particle fluctuations in Gaussian states}\label{sec:gauss-evo}

As mentioned, the QSSEP introduces simplifications compared to generic RUC dynamics, as the Hamiltonian generator is quadratic in the fermions. Therefore, we may apply techniques of Gaussian fermionic states to compute the bipartite entanglement and the moment generating functions. We recall that fermionc Gaussian states are defined by the fact that they satisfy Wick theorem~\cite{bravyi2004lagrangian,surace2022fermionic}. As a consequence, the density matrix of a Gaussian state $\ket{\psi}$ is completely specified by its correlation (or covariance) matrix 
\begin{equation}
\Gamma_{i,j}= \langle\psi| \hat c^\dagger_i \hat c_j|\psi \rangle\,.
\end{equation}

Importantly, the initial state~\eqref{eq:DW_state} is Gaussian, its correlation matrix reading
\be\label{eq:initial_covariance_matrix}
\Gamma(0)=
\begin{pmatrix}
	\id_{M,M} & 0_{M,L-M}\\
	0_{L-M,M} & 0_{L-M,L-M}
\end{pmatrix}\,,
\ee
where pedices indicate the dimensions of the matrices. The evolved state $\ket{\Psi(t)}$ thus remains Gaussian at all times, and one can express the corresponding covariance matrix as
\be\label{eq:gamma_function}
\Gamma(t)=U(t)\Gamma(0)U^{\dagger}(t)\,.
\ee
Here, $U(t)$ is the single-particle unitary matrix associated with the evolution operator $\mathcal{U}(t)$ defined by Eq.~\eqref{eq:unitary_ev}, cf. Sec.~\ref{sec:jacobi_process}.

The covariance matrix fully determines the properties of the state, including the value of the R\'enyi entropies and the full-counting statistics. Indeed, denoting by $A^{(\ell)}$ the matrix obtained by selecting the first $\ell$ rows and columns of a matrix $A$, we have~\cite{vidal2003entanglement} 
\be
S_\ell^{(q)}(t)=(1-q)^{-1}\sum_{j=1}^\ell\ln\left[\lambda_j(t)^q+(1-\lambda_j(t))^q\right]\,,
\ee
where $\{\lambda_j(t)\}_{j=1}^\ell$ are the eigenvalues of $\Gamma^{(\ell)}(t)$ (which satisfy $0\leq \lambda_j(t)\leq 1$). 
Similarly (see e.g. Ref.~\cite{bernard2025large})
\begin{align}
Z_\ell(\phi,t)=\langle e^{i \phi \hat N_\ell}\rangle= \det\left[ {\bm 1}_{\ell\times \ell} + \left(e^{i \phi}-{\bm 1}_{\ell\times \ell}\right) \Gamma^{(\ell)}(t)\right]\nn
=\prod_{j=1}^M \left(1+ \big(e^{i \phi}-1\big)\lambda_j(t)\right).
\end{align}
These formulas allow us to study numerically the evolution of the entanglement entropy and full-counting statistics efficiently, as they require a computational cost scaling only polynomially, rather than exponentially, with the system size $L$. However, they are in general not suitable for analytic investigations, because for general connectivity it is hard to determine the eigenvalues $\lambda_j(t)$. For this reason, previous approaches to entanglement evolution in the local and non-local QSSEP relied on replica calculations~\cite{swann2025spacetime,tiutiakina2025field}. In the next section, instead, we will show that in our quench protocol recent RMT results allow us to write down a closed system of stochastic differential equations for $\lambda_j(t)$, arriving at exact and explicit results for entanglement and particle fluctuations.

\section{The Jacobi process}
\label{sec:jacobi_process}

\subsection{Stochastic differential equations for the correlation-matrix eigenvalues}

The quadratic generator~\eqref{eq:dH} induces a stochastic evolution for the covariance matrix $\Gamma(t)$, which is driven by a unitary operator $U(t)$, cf. Eq.~\eqref{eq:gamma_function}. The relation between $\mathcal{U}(t)$ in Eq.~\eqref{eq:unitary_ev} and $U(t)$ in Eq.~\eqref{eq:gamma_function} follows from the theory of fermionic Gaussian states~\cite{bravyi2004lagrangian}. In particular, it is straightforward to derive the differential equation satisfied by the matrix $U(t)$, reading
\be\label{eq:SDE_U}
dU(t)=-idW_tU(t)-\frac{1}{2}dW_tdW_tU(t)=-idW_tU(t)-\frac{1}{2} U(t)dt\,,
\ee
where we used~\eqref{eq:unitarity_w}, while $dW_t$ is the matrix introduced in Eq.~\eqref{eq:dw_mat}.

Now, writing
\be\label{eq:U_block}
U(t)=
\begin{pmatrix}
	X_{\ell,M}(t)& \tilde{Y}_{\ell,L-M}(t)\\
	Y_{L-\ell,M} (t)& \tilde{X}_{L-\ell,L-M}(t)
\end{pmatrix}\,,
\ee
and considering the form of the initial covariance matrix~\eqref{eq:initial_covariance_matrix}, we immediately obtain the covariance matrix restricted to $A_\ell$ at time $t$, namely
\be
\Gamma_{\ell}(t)=X(t)X^{\dagger}(t)\,,
\ee
which is a $\ell\times \ell$ matrix. We note that $\Gamma_{\ell}(t)$ has the same non-zero eigenvalues of the $M\times M$ matrix
\be
J(t)=X^{\dagger}(t)X(t)\,.
\ee
Indeed, for any matrix $A$, if $\lambda$ is a non-zero eigenvalue of $AA^{\dagger}$, i.e. $\exists \ket{v}$ s.t. $AA^{\dagger}\ket{v}=\lambda \ket{v}$, then, defining $\ket{w}=A^{\dagger}\ket{v}$ we have
\be
A^{\dagger}A \ket{w}=A^{\dagger}(A A^{\dagger})\ket{v}=\lambda\ket{w}\,,
\ee
so $\lambda$ is a non-zero eigenvalue of $A^\dagger A$ (note that $\ket{w}\neq 0$, because $A\ket{w}=AA^\dagger \ket{v}=\lambda\ket{v}\neq 0$).

At finite time $t$, all non-zero eigenvalues of $J(t)$  and $\Gamma_\ell(t)$ are expected to be non-degenerate with probability $1$. Therefore, in order to study the distribution of the eigenvalues of $\Gamma_\ell(t)$, we may study the eigenvalues of $J(t)$. This observation is particularly useful, because the dynamics of $J(t)$ has been studied in the context of the so-called Jacobi process, introduced in Ref.~\cite{doumerc2005matrices}. In the following, we adapt the derivation of~\cite{doumerc2005matrices} to our framework, to write down a closed SDE for the matrix $J(t)$.

Let us first write $dW_t$ in block diagonal form as
\be
dW_t=
\begin{pmatrix}
	dA_{\ell,\ell}& dB_{\ell, L-\ell}\\
	dB^\dagger_{\ell, L-\ell} & dC_{L-\ell,L-\ell}
\end{pmatrix}\,.
\ee
Using the decomposition~\eqref{eq:U_block}, it follows from~\eqref{eq:SDE_U} that
\begin{align}
dX(t)&=-i(dA)X(t)-i(dB)Y(t)-\alpha_L X(t)dt\,,\\
dX^\dagger(t)&=iX^\dagger(t)(dA)+iY^\dagger(t)(dB^\dagger)-\alpha_L X^{\dagger}(t)dt\,,
\end{align}
where $\alpha_L=(L-1)/(2L)$ and we used that $dA$ is Hermitian. In order to derive a differential equation for $J(t)$, we use
\begin{align}
(dA)^\dagger(dA)&=\frac{\ell}{L}\id_{\ell,\ell}dt,\\
(dB)^\dagger(dB)&=\frac{\ell}{L}\id_{L-\ell,L-\ell}dt,
\end{align}
and also
\be
Y^\dagger(t)Y(t)=\id-X^\dagger(t)X(t)\,,
\ee
which follows from unitarity. From the equations above, making use of It\^{o} rules and some algebra, we obtain
\begin{align}
dJ(t)=d[X^\dagger(t)X(t)]&=(dX^\dagger(t))X(t)+X^\dagger(t)dX(t)+dX^\dagger(t)dX(t)\\
&=i Y^\dagger(t) (dB)^\dagger X(t)-i X^\dagger(t) (dB)Y(t)+\left(\frac{\ell}{L}\id-J\right) dt\,.
\end{align}
Defining now
\be
d\mathcal{W}(t)=J^{-1/2}(t)X^\dagger (t)dBY(t)[\id-J(t)]^{-1/2}\,,
\ee
the SDE becomes
\begin{align}\label{eq:jacobi_process}
	dJ(t)&=[\id-J(t)]^{1/2}d\mathcal{W}^\dagger(t)J^{1/2}(t)+J^{1/2}(t)d\mathcal{W}(t)[\id-J(t)]^{1/2}+\left(\frac{\ell}{L}\id-J\right) dt\,.
\end{align}
Finally, by direct computation, we can show that $d\mathcal{W}(t)$ are complex Brownian motions, i.e.
\begin{align}
d \mathcal{W}_{x,y} d \bar{\mathcal{W}}_{t,z}  = \frac{1}{L}\delta_{x,t}\delta_{y,z}\,.
\end{align}

The SDE~\eqref{eq:jacobi_process} defines a \emph{Jacobi process}. In this case, it follows from the so-called Bru’s Theorem~\cite{doumerc2005matrices,deleaval2018moments} that the eigenvalues of $J(t)$ also satisfy a SDE
\begin{align}\label{eq:lambda-jacobi}
d \lambda_{i}(t)=& \sqrt{\frac{2}{L}}\sqrt{\left(\lambda_{i}(t)\left(1-\lambda_{i}(t)\right)\right.} d \nu_{i}(t)\nonumber\\[4pt]
 &+\frac{1}{L}\left[\ell-L \lambda_{i}(t)+\sum_{j \neq i} \frac{\lambda_{i}(t)\left(1-\lambda_{j}(t)\right)+\lambda_{j}(t)\left(1-\lambda_{i}(t)\right)}{\lambda_{i}(t)-\lambda_{j}(t)}\right] d t\,,
 \end{align}
where ${\nu_j(t)}_{j=1}^M$ are real Brownian motions
\be
d\nu_i(t) d\nu_k(t)=\delta_{i,k}\,dt.
\ee 

\subsection{The thermodynamic limit}

The closed system of stochastic differential equations~\eqref{eq:lambda-jacobi} can be in principle solved numerically, yielding a complete characterization of any property that only depends on the spectrum of the reduced covariance matrix, including the bipartite entanglement and the full counting statistics. In fact, it is possible to go further, and exploit the underlying RMT structures to obtain an analytic expression of all the eigenvalue moments in the thermodynamics limit.

To be precise, let us consider the limit where $M, L, \ell\to\infty$, keeping the ratios
 \be
\theta:=\frac{\ell}{L}\,,\qquad \zeta:=\frac{M}{\ell}\,,\qquad \frac{M}{L}=\theta \zeta\,,
 \ee
 fixed. We shall denote this limit as $\lim_{\rm Th}$ in the expressions below. It was proven in the mathematical literature that in this limit the expectation value of the normalized trace of any finite tuple of matrices drawn from $J_{t}$ converges~\cite{demni2008free,collins2018spectral}. In particular, the limits
 \be
 m_{n}^{ \zeta, \theta}(t):=\lim _{\rm Th} \frac{1}{M} \mathbb{E}\left(\operatorname{tr}\left[\left(J_{t }\right)^{n}\right]\right)  
\ee 
exist for all $n>0$ and $t>0$.  In fact, in this case the problem can  be formulated in terms of an abstract probability space, and the Jacobi process converges to the so-called \emph{free Jacobi process}~\cite{demni2008free}, see also Refs.~\cite{demni2008free,demni2018inverse,collins2018spectral,deleaval2018moments,demni2020hermitian}. The goal of this section is to show how some of the results derived in these works  can be exploited directly for our purposes,  leading to a  fully analytic expression for the  entanglement dynamics. 

First of all, introducing the  resolvent and its normalized trace
\be
\mathcal{R}(z):=\frac{1}{J_t-z}, \qquad R(z):=\frac{1}{M}\operatorname{tr} \mathcal{R}(z)\,,
\ee
we note that in the thermodynamic  limit
\be
\lim_{\rm Th} R(z)=\frac{1}{z} \sum_{n=0}^{\infty} \frac{m_{n}(t)}{z^{n}}\,,
\ee
where we omitted explicit  dependence on $\theta$, $\zeta$.  Now, starting from~\eqref{eq:jacobi_process}, a key  result of Refs.~\cite{demni2008free,demni2012spectral} was a differential equation for the moments $m_n(t)$,  which also allows one to derive a differential equation for the trace of the resolvent $R(z)$. Specifically, one can  derive
\be\label{eq:recurrence_equation}
\partial_{t} m_{n}(t)=-n m_{n}(t)+\theta n m_{n-1}(t)+\zeta \theta n \sum_{k=0}^{n-2} m_{n-k-1}(t)\left(m_{k}(t)-m_{k+1}(t)\right)\,,
\ee
from which it follows
\be
\partial_{t} R_{t}=\partial_{z}\left\{[(1-2 \zeta \theta) z-\theta(1-\zeta)] R_{t}+\zeta \theta z(z-1) R_{t}^{2}\right\}\,.
\ee
These two simple equations are not  easy to derive, and  involve sophisticated mathematical tools. In the following, we will simply exploit their consequences for our problem.\\

A final crucial result is that Eq.~\eqref{eq:recurrence_equation} can be solved explicitly: the solution can be found in Ref.~\cite{demni2008free,demni2012spectral} for $\theta=1/2$, $\zeta=1$ and in Ref.~\cite{demni2018inverse} for arbitrary $\theta$ and $\zeta$. Quite surprisingly, the form of the solution is very simple for $\theta=1/2$, $\zeta=1$, while it is quite involved in the latter case. For simplicity, in the following we will restrict to $\theta=1/2$, $\zeta=1$. Physically, this corresponds to a  domain wall initial state with a bipartition of the system into two halves. 

\section{Physical applications}\label{sec:phys-app}

\subsection{The entanglement dynamics}\label{sec:entanglement}

We now discuss our first application. We begin by reporting the exact solution to Eq.~\eqref{eq:recurrence_equation} with $\theta=1/2$ and $\zeta=1$, which reads~\cite{demni2012spectral}
\be\label{eq:explicit_moment}
m_{n}(t)=\frac{1}{2^{2 n}}\left(\begin{array}{c}2 n \\ n\end{array}\right)+\frac{1}{2^{2 n-1}} \sum_{k=1}^{n}\left(\begin{array}{c}2 n \\ n-k\end{array}\right) \frac{1}{k} L_{k-1}^{(1)}(2 k t) e^{-k t}\,,
\ee
where $L^{(1)}_n(x)$ is the $n$-th Laguerre polynomial of index $1$. Let us show how this result can be used to arrive at a fully analytical expression for the entanglement dynamics in the thermodynamic limit (when $\ell/M=1$ and $\ell/L=1/2)$. First, we use the expansion
\be\label{eq:expansion}
-x\ln x-(1-x)\ln (1-x)=\ln 2-\sum_{n=1}^\infty\frac{ 2^{2n}}{(-1 +2 n) 2n}\left(x-\frac{1}{2}\right)^{2n}\,,
\ee
which is convergent in the interval $x\in (0,1)$. Next, we recall that the normalized von Neumann entanglement entropy can be written as
\be\label{eq:s_lambdas}
s(t)=-\frac{1}{\ell}\sum_{j=1}^\ell\left[\lambda_j\ln\lambda_j+ (1+\lambda_j)\ln(1+\lambda_j)\right]\,,
\ee
where $\lambda_j$ are the eigenvalues of $J(t)$. From~\eqref{eq:s_lambdas}, taking the expectation value in the thermodynamic limit and using Eqs.~\eqref{eq:explicit_moment} and~\eqref{eq:expansion}, we obtain
\be
s(t)=\log (2)-\sum _{n=1}^{\infty } \sum _{k=0}^{2 n} \frac{(-1)^k 2^{k-1} \binom{2 n}{k}
	\left(2^{1-2 k} \sum _{k=1}^k \frac{e^{-k t} L_{k-1}^{(1)}(2 k t)}{k}+2^{-2 k} \binom{2
		k}{k}\right)}{n (2 n-1)}\,.
\ee
\begin{figure}[t]
\centering
\includegraphics[width=0.7\textwidth]{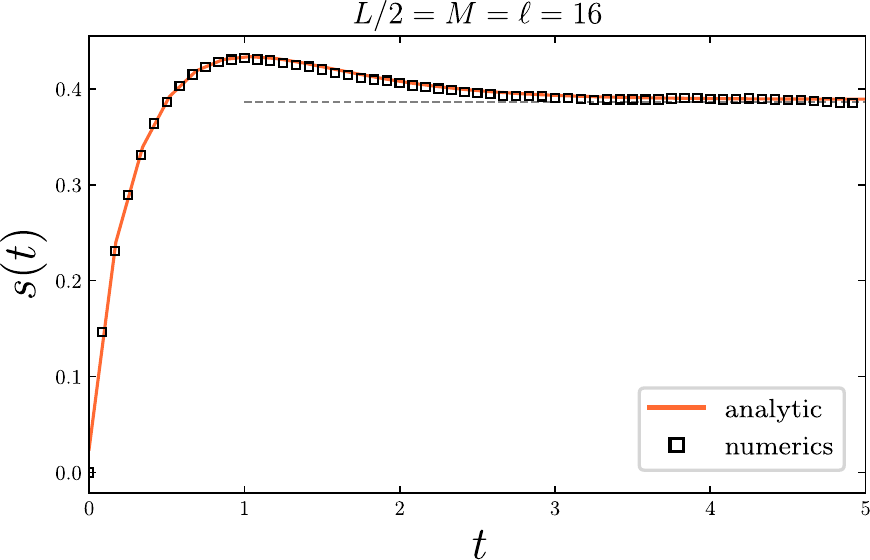}
\caption{Entropy density dynamics from a domain-wall initial state with $L=32$ and $M=\ell=L/2$. The solid line corresponds to the analytic prediction in Eq.~\eqref{eq:final}, while the symbols are obtained from numerical simulations of the quantum dynamics (see Appendix~\ref{app:num}), averaged over 200 realizations. The dashed horizontal line marks the steady-state value $s(\infty)=2\log(2)-1$.}\label{fig:comparison-entanglement}
\end{figure}
This expression can be simplified a bit, yielding the final result
\begin{align}\label{eq:final}
	s(t)=&2\log(2)-1-\sum_{n=1}^\infty\frac{1}{n(2n-1)}\sum_{k=1}^n\frac{(-2)^{-2k}}{2k}\binom{2n}{2k}
	\nonumber \\[4pt]
	&  \; \times\ _2F_1\left(2 k+\frac{1}{2},2 k-2 n;4 k+1;2\right)e^{-k t} L_{k-1}^{(1)}(2 k t)
\end{align}
where $_2F_1$ is the Gaussian hypergeometric function. We note that the long-time limit of Eq.~\eqref{eq:final} can be obtained from the moment formula~\eqref{eq:explicit_moment}. For any fixed $n$, the Laguerre contribution is a finite sum over $k$ multiplied by $e^{-kt}$, and therefore vanishes as $t\to\infty$. This yields the stationary moments
\be
m_n(\infty)=\frac{1}{2^{2n}}\binom{2n}{n},
\ee
which correspond to the arcsine eigenvalue distribution $\rho_\infty(\lambda)=\big[\pi\sqrt{\lambda(1-\lambda)}\big]^{-1}$ on $[0,1]$.  This distribution is the universal spectral measure of the correlation matrix in the stationary state~\cite{liu2018quantum,bernard2021entanglement}, reflecting the fact that the dynamics generates an effectively random Gaussian state subject to particle-number conservation. As a result, the steady state is maximally mixed within this constrained class, but not maximally entangled. This is precisely why the entropy density saturates at
\be
s(\infty)=2\log(2)-1<\log(2),
\ee
in agreement with the constant term in Eq.~\eqref{eq:final}. We have tested the validity of Eq.~\eqref{eq:final} against numerical simulations for finite system sizes. An example of our numerical data is reported in Fig.~\ref{fig:comparison-entanglement}, showing excellent agreement already at moderate values of $L$. Details of the numerical simulations can be found in Appendix~\ref{app:num}.
\subsection{Charge full-counting statistics}
\begin{figure}[t]
\centering
\includegraphics[width=\textwidth]{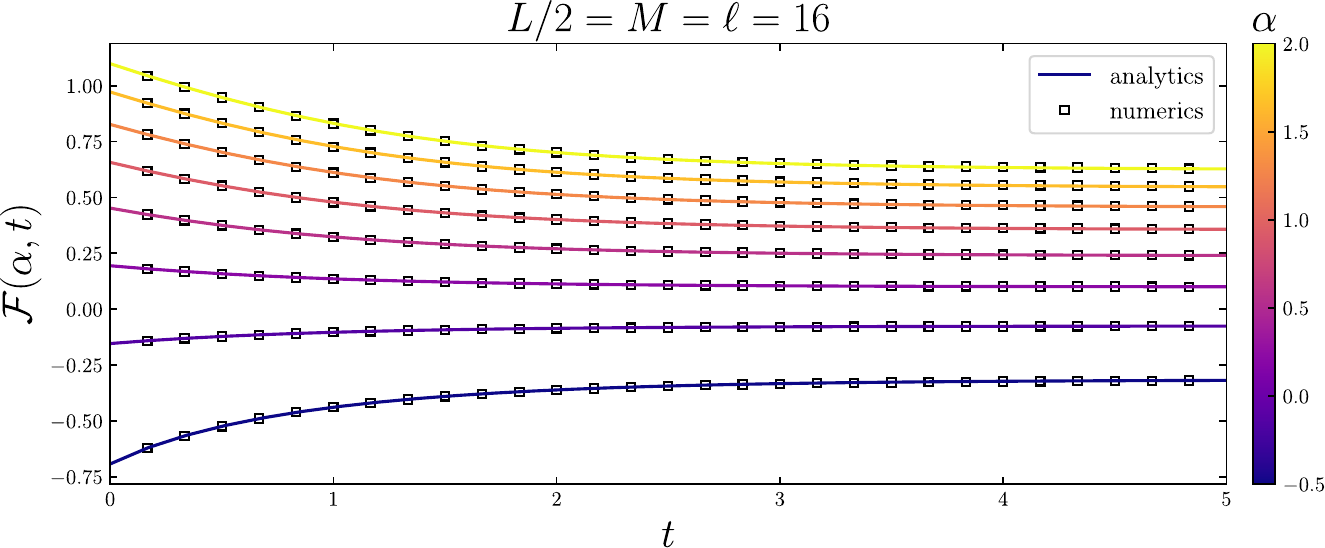}
\caption{Charge full-counting statistics dynamics from a domain-wall initial state with $L=32$ and $M=\ell=L/2$. The solid lines correspond to the analytic predictions in Eq.~\eqref{eq:fcs-quantum}, while the symbols are obtained from numerical simulations of the quantum dynamics (see Appendix~\ref{app:num}), averaged over 200 realizations. Several values of $\alpha$ are shown, as indicated by the color legend.}\label{fig:fcs-comparison}
\end{figure}

We also consider the cumulant generating function in the thermodynamic limit, defined as
\begin{equation}
{\cal F}(\phi,t)=\lim_{\mathrm{Th}}\frac{1}{\ell}\,\mathbb{E}\big[\log Z_\ell(\phi,t)\big].
\end{equation}
Expanding the logarithm as
\begin{equation}
\log\bigl(1+(e^{i\phi}-1)\lambda\bigr)
=\sum_{n\ge1}\frac{(-1)^{n+1}}{n}(e^{i\phi}-1)^n\lambda^n,
\end{equation}
one obtains the explicit expression
\begin{equation}\label{eq:fcs-quantum-moments}
{\cal F}(\phi,t)=\sum_{n\ge1}\frac{(-1)^{n+1}}{n}(e^{i\phi}-1)^n\, m_n(t)
\end{equation}
in terms of the moments $m_n(t)$ introduced above. In the domain-wall configuration $M=\ell=L/2$, inserting the explicit expression \eqref{eq:explicit_moment} for the moments yields a closed form for the generating function. Writing $\alpha:=e^{i\phi}-1$, after simple algebra detailed in Appendix~\ref{app:algebra-fcs}, one finds
\begin{align}\label{eq:fcs-quantum}
{\cal F}(\alpha,t)=&{\cal F}_{\rm ss}(\alpha)-2\sum_{k\ge1} e^{-kt} L^{(1)}_{k-1}(2kt)
\left(-\frac{\alpha}{4}\right)^k
{}_2F_1\!\left(k+\tfrac12,k;2k+1;\alpha\right),
\end{align}
with stationary value,
\be\label{eq:fcs-quantum-ss}
{\cal F}_{\rm ss}(\alpha)=2\log\!\left(\frac{1+\sqrt{1+\alpha}}{2}\right).
\ee
This expression is exact in the thermodynamic limit. In Fig.~\ref{fig:fcs-comparison}, a comparison with numerical simulations at finite size shows excellent agreement already for modest system sizes $L\lesssim 32$. \\

We emphasize that the exact knowledge of the moments $m_n(t)$ of the eigenvalues $\lambda_i(t)$, governed by the stochastic evolution~\eqref{eq:lambda-jacobi}, allows one to characterize the thermodynamic-limit behavior of generic spectral observables of the form
\be\label{eq:X-spectral}
X_f(t):=\frac{1}{\ell}\sum_{i=1}^\ell f(\lambda_i(t)),
\ee
under mild smoothness assumptions on the function $f(\lambda)$. This class includes, in particular, the entanglement entropy and the charge full-counting statistics.  

In Appendix~\ref{app:self-avg}, we show that such observables (in particular $s(t)$ and ${\cal F}(\alpha,t)$) exhibit a self-averaging behavior,
\be\label{eq:self-avg}
\Var\big(X_f(t)\big) = \E\Big[\big(X_f(t)-\E[X_f(t)]\big)^2\Big]
= {\cal O}\!\left(\frac{1}{L\ell}\right).
\ee
This property will play a key role in comparing the charge fluctuation statistics of the fully-connected QSSEP with that of the classical process associated with the noise-averaged dynamics, discussed in the following sections.
\section{Comparison to the classical SSEP}\label{sec:classical-ssep}

\subsection{Noise-averaged dynamics: all-to-all SSEP}

We now consider the noise-averaged density matrix $\bar\rho(t):=\E[\hat\rho(t)]$, which evolves according to
\be
\partial_t \bar\rho(t) = \mathcal L\big(\bar\rho(t)\big).
\ee
Expanding $\bar\rho(t)$ in the occupation-number basis at fixed particle number $M$,
\be
\bar\rho(t)=\sum_{\eta,\eta'} \rho_{\eta,\eta'}(t)\,|\eta\rangle\langle\eta'|,
\qquad
\eta\in\{0,1\}^L,\quad \sum_i \eta_i=M,
\ee
we decompose it into diagonal and off-diagonal parts,
\be
\bar\rho(t)=\bar\rho_{\mathrm{diag}}(t)+\bar\rho_{\mathrm{off}}(t),
\ee
with $\bar\rho_{\mathrm{diag}}(t)=\sum_\eta P_t(\eta)|\eta\rangle\langle\eta|$. A direct computation using fermionic exclusion shows that the diagonal sector is invariant under $\mathcal L$, and evolves according to
\be\label{eq:SSEP-master}
\partial_t P_t(\eta)
=
\frac{1}{L}\sum_{j\neq k}
\Big[
\eta_j(1-\eta_k) P_t(\eta^{j\to k})
-
\eta_k(1-\eta_j) P_t(\eta)
\Big],
\ee
with $\eta^{j\to k}$ denoting the configuration obtained from $\eta$ after a particle hopping from $j\to k$.\\
This is precisely the generator of the symmetric simple exclusion process on a complete graph~\cite{derrida2007non,Mendona2013}, where particles hop from any occupied site $k$ to any empty site $j$ with rate $1/L$.

Similarly, the off-diagonal sector is preserved by $\mathcal L$, and evolves independently under dephasing. Therefore, for observables depending only on site occupations (in particular the subsystem particle number $\hat N_\ell$) the noise-averaged dynamics is exactly described by the all-to-all SSEP.

\subsection{Stationary charge full-counting statistics}

We consider the all-to-all SSEP with $L$ sites and $M$ particles, and the subsystem particle number
\be
N_\ell=\sum_{j\in A_\ell}\eta_j.
\ee
The induced dynamics for $N_\ell$ is a birth-death process on $n=0,\dots,\min(M,\ell)$ with rates
\be\label{eq:birthdeath_rates}
b_n=\frac{(\ell-n)(M-n)}{L},
\qquad
d_n=\frac{n\bigl(L-\ell-M+n\bigr)}{L}.
\ee
The stationary measure $\pi(n)$ follows from the detailed balance condition
\be\label{eq:detailed-bal}
\pi(n)\,b_n=\pi(n+1)\,d_{n+1},
\ee
which, upon iteration, yields the hypergeometric distribution
\be
\pi(n)=\frac{\binom{\ell}{n}\binom{L-\ell}{M-n}}{\binom{L}{M}}.
\ee
In the case $M=\ell=L/2$, this reduces to
\be
\pi(n)=\binom{2\ell}{\ell}^{-1}\binom{\ell}{n}^2.
\ee
Hence, the stationary generating function for the classical all-to-all SSEP 
\be
Z^{\rm class}_{{\rm ss};\ell}(\alpha):=\sum_{n=0}^{M}\pi(n)(1+\alpha)^n,
\qquad
{\cal F}^{\rm class}_{\rm ss}(\alpha):=
\lim_{L\to\infty}\frac{1}{M}\log Z^{\rm class}_{{\rm ss};\ell}(\alpha),
\ee
read (for $M=\ell=L/2$)
\be
Z^{\rm class}_{{\rm ss};\ell}(\alpha)
=
\frac{1}{\binom{2\ell}{\ell}}
\sum_{n=0}^{\ell}\binom{\ell}{n}^2 (1+\alpha)^n.
\ee
Note that $Z^{\rm class}_{{\rm ss};\ell}(\alpha)$ is independent of the connectivity, and has the same form for the local QSSEP.\\

The large-$L$ asymptotics is obtained by a saddle-point evaluation. Writing $n=x\ell$ and using Stirling’s approximation,
\be
\binom{\ell}{x\ell}\asymp e^{\ell h(x)},
\quad
h(x):=-x\log x-(1-x)\log(1-x),
\ee
one finds
\be
\log Z^{\rm class}_{{\rm ss};\ell}(\alpha)
\sim
\ell \sup_{x\in[0,1]}
\Big[2h(x)+x\log(1+\alpha)-2\log 2\Big].
\ee
Extremization yields $x_*=\frac{s}{1+s}$ with $s=\sqrt{1+\alpha}$, and evaluating the exponent gives
\be\label{eq:fcs-ss}
{\cal F}^{\rm class}_{\rm ss}(\alpha)
={\cal F}_{\rm ss}(\alpha)=
2\log\!\left(\frac{1+\sqrt{1+\alpha}}{2}\right).
\ee
We thus observe that, in the thermodynamic limit, the stationary charge full-counting statistics of the classical process coincides with that of the all-to-all QSSEP, cf.~Eq.~\eqref{eq:fcs-quantum-ss}, in agreement with the results of Refs.~\cite{bauer2024bernoulli,bernard2025large,albert2026universal,costa2025emergence} for the boundary-driven case. Going beyond previous work, however, we can now ask whether such equivalence is restricted to the steady state, or extends to finite times.
\\
At lowest orders, a direct calculation (see Appendix~\ref{app:low-cum-ssep}) shows that both the mean and the variance of the all-to-all SSEP agree with their quantum counterparts already at finite times. Motivated by this observation, we show in the following section that the equations of motion for the full fluctuation statistics, as encoded in the classical and quantum generating functions, coincide up to finite-size corrections.
\begin{figure}[t]
\centering
\includegraphics[width=.8\textwidth]{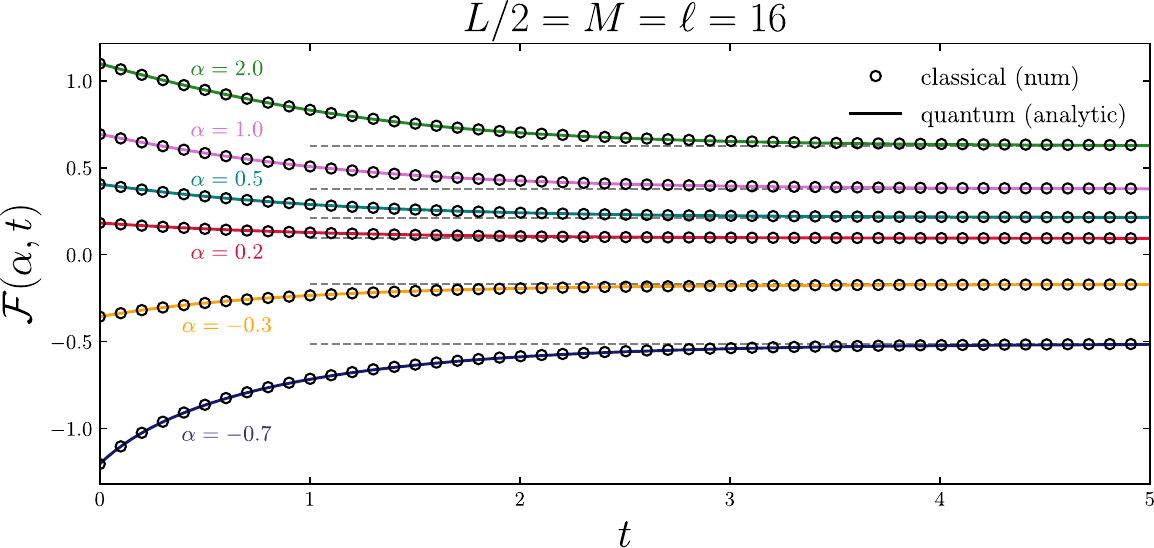}
\caption{Comparison between the classical and quantum dynamics of the charge full-counting statistics. \emph{Symbols}:~numerical data for the time evolution of ${\cal F}^{\rm class}_\ell(\alpha,t)$ for $\ell=M=L/2=16$ as a function of time for different values of $\alpha$ (see color legend), obtained from Eq.~\eqref{eq:P-formal} by direct matrix exponentiation. \emph{Solid lines}:~analytic solution for the quantum dynamics ${\cal F}(\alpha,t)$, given in Eq.~\eqref{eq:fcs-quantum}. Dashed horizontal lines indicate the steady-state value in Eq.~\eqref{eq:fcs-ss}.}
\end{figure}
\section{Finite-time dynamics: quantum and classical charge fluctuations}\label{sec:finite-time-comparison}

\subsection{Quantum evolution equation}

We start by considering the quantum process introduced in Sec.~\ref{sec:the_model}. We define the following finite-size cumulant generating function for $M=\ell$
\be
{\cal F}^{\rm qu}_\ell(\alpha,t):=\frac{1}{\ell}\log Z_\ell(\alpha,t)
=\frac{1}{\ell}\sum_{i=1}^{\ell}\log\big(1+\alpha\lambda_i(t)\big),
\ee
such that ${\cal F}(\alpha,t)=\lim_{\rm Th}\E[{\cal F}^{\rm qu}_\ell(\alpha,t)]$. We will work with ${\cal F}^{\rm qu}_\ell$ as defined above, since it is important to keep track of finite-size corrections in the derivation below. Using It\=o calculus together with the SDE \eqref{eq:lambda-jacobi} for $\lambda_i(t)$, one finds the structure
\be\label{eq:tmp1}
\partial_t \E[{\cal F}^{\rm qu}_\ell(\alpha,t)] = \E[\tilde {A}_\alpha(t)],
\ee
where the drift $\tilde {A}_\alpha$ can be computed explicitly from Eq.~\eqref{eq:lambda-jacobi}. After symmetrizing the interaction term and exploiting cancellations, one obtains the {\it exact} equation
\be\label{eq:tmp2}
\partial_t \E[{\cal F}^{\rm qu}_\ell(\alpha,t)]
=
\frac{1}{\ell L}\E\!\left[\sum_{i=1}^{\ell}
\frac{\alpha(\ell-L\lambda_i)}{1+\alpha\lambda_i}\right]
-\frac{\alpha^2}{\ell L}
\E\!\left[
\left(\sum_{i=1}^{\ell}\frac{\lambda_i}{1+\alpha\lambda_i}\right)
\left(\sum_{j=1}^{\ell}\frac{1-\lambda_j}{1+\alpha\lambda_j}\right)
\right].
\ee
The details of the derivation of Eqs.~\eqref{eq:tmp1} and \eqref{eq:tmp2} can be found in Appendix~\ref{app:algebra2}. Introducing the biased density
\be
n_\alpha(t):=\partial_\alpha {\cal F}^{\rm qu}_\ell(\alpha,t)
=\frac{1}{\ell}\sum_{i=1}^{\ell}\frac{\lambda_i(t)}{1+\alpha\lambda_i(t)},
\ee
the evolution equation can be rewritten as
\be
\partial_t \E[{\cal F}^{\rm qu}_\ell(\alpha,t)]
=
\frac{\alpha\ell}{L}
-\alpha\left(1+\frac{2\alpha\ell}{L}\right)\E[n_\alpha]
+\frac{\alpha^2\ell}{L}(1+\alpha)\E[n_\alpha^2].
\ee
In the case $\ell=L/2$, this simplifies to
\be\label{eq:Xqu}
\partial_t \E[{\cal F}^{\rm qu}_\ell(\alpha,t)]
=
\frac{\alpha}{2}
-\alpha(1+\alpha)\E[n_\alpha]
+\frac{\alpha^2}{2}(1+\alpha)\E[n_\alpha^2].
\ee

To close the equation, we use self-averaging. Since $n_\alpha$ is itself a spectral observable, its fluctuations are suppressed as $\Var(n_\alpha)=\mathcal O(L^{-2})$ (see Eq.~\eqref{eq:self-avg} and Appendix~\ref{app:self-avg}), implying
\be
\E[n_\alpha^2]=\E[n_\alpha]^2+\mathcal O(L^{-2}).
\ee
We conclude therefore that the charge full-counting statistics for the all-to-all QSSEP for $M=\ell=L/2$ satisfies the equation
\be
\partial_t \E[{\cal F}^{\rm qu}_\ell(\alpha,t)]
=
\frac{\alpha}{2}
-\alpha(1+\alpha)\E[n_\alpha]
+\frac{\alpha^2}{2}(1+\alpha)\E[n_\alpha]^2
+\mathcal O(L^{-2}).
\ee
The initial condition follows from $\lambda_i(0)=1$, yielding ${\cal F}^{\rm qu}_\ell(\alpha,0)=\log(1+\alpha)$. By construction, the solution to this equation is given in Eq.~\eqref{eq:fcs-quantum}, up to finite-size corrections.
\subsection{Classical evolution equation}

We now consider the classical process. We introduce the moment generating function of the subsystem occupation number,
\be
Z^{\rm class}_\ell(s,t):=\E\big[s^{N_\ell(t)}\big]
=
\sum_{n=0}^{\min(M,\ell)} P_n(t)\, s^n,
\qquad s=1+\alpha,
\ee
where $P_n(t)={\rm Prob}(N_\ell(t)=n)$. The dynamics of $P_n(t)$ is governed by the birth–death equation
\be\label{eq:Markov-birth-death}
\partial_t P_n
=
b_{n-1}P_{n-1}+d_{n+1}P_{n+1}-(b_n+d_n)P_n,
\ee
with rates given in Eq.~\eqref{eq:birthdeath_rates}. In the domain-wall setting $M=\ell=L/2$, these simplify to
\be
b_n=\frac{(\ell-n)^2}{L},
\qquad
d_n=\frac{n^2}{L}.
\ee
This defines a finite-dimensional Markov process with support $n\in[0,\ell]$, and boundary conditions encoded in $d_0=0$ and $b_\ell=0$. The initial condition reads $P_n(0)=\delta_{n,\ell}$. Equivalently, introducing the probability vector ${\bm P}(t)=(P_0(t),\dots,P_\ell(t))^{\mathsf T}$, the dynamics can be written in matrix form as
\be\label{eq:P-formal} 
\partial_t \mathbf P(t)=\mathbb G\,\mathbf P(t),
\ee
where $\mathbb G$ is a tridiagonal Markov generator with entries
\be
(\mathbb G)_{n,n}=-(b_n+d_n),\qquad
(\mathbb G)_{n,n-1}=b_{n-1},\qquad
(\mathbb G)_{n,n+1}=d_{n+1}.
\ee
The formal solution reads $\mathbf P(t)=e^{t\mathbb G}\mathbf P(0)$, which allows for straightforward numerical evaluation via direct matrix exponentiation for moderate system sizes.\\

To make progress, it is convenient to introduce the finite-size cumulant generating function for $M=\ell$
\be
{\cal F}_\ell^{\rm class}(\alpha,t)
:=\frac{1}{\ell}\log Z^{\rm class}_\ell(1+\alpha,t),
\ee
so that ${\cal F}^{\rm class}=\lim_{\rm Th} {\cal F}^{\rm class}_\ell$. Differentiating with respect to time and using Eq.~\eqref{eq:Markov-birth-death}, one finds
\be
\partial_t {\cal F}_\ell^{\rm class}(\alpha,t)
=
\frac{1}{\ell}
\left\langle
\alpha\, b_n - \frac{\alpha}{1+\alpha}\, d_n
\right\rangle_{{\rm class},\alpha},
\ee
where $\langle\cdot\rangle_{{\rm class},\alpha}$ denotes expectation with respect to the tilted measure proportional to $P_n(t)(1+\alpha)^n$. Introducing the biased density
\be
n_\alpha^{\rm class}(t):=\partial_\alpha {\cal F}_\ell^{\rm class}(\alpha,t),
\ee
one can express moments of $n$ in terms of derivatives with respect to $\alpha$. After straightforward algebra, this yields the {\it exact} evolution equation
\begin{align}
\partial_t {\cal F}_\ell^{\rm class}(\alpha,t)
=&
\frac{\alpha\ell}{L}
-\alpha\left(1+\frac{2\alpha\ell}{L}\right)n_\alpha^{\rm class}
+\frac{\alpha^2\ell}{L}(1+\alpha)(n_\alpha^{\rm class})^2\nonumber\\[4pt]
&+\frac{\alpha(\alpha-1)}{L}n_\alpha^{\rm class}
+\frac{\alpha(\alpha-1)(1+\alpha)}{L}\partial_\alpha^2 {\cal F}_\ell^{\rm class}(\alpha,t).
\end{align}

In the case $\ell=L/2$, this simplifies to
\be\label{eq:Xclass}
\partial_t {\cal F}_\ell^{\rm class}(\alpha,t)
=
\frac{\alpha}{2}
-\alpha(1+\alpha)n_\alpha^{\rm class}
+\frac{\alpha^2}{2}(1+\alpha)(n_\alpha^{\rm class})^2
+\mathcal O(L^{-1}),
\ee
with initial condition ${\cal F}_\ell^{\rm class}(\alpha,0)=\log(1+\alpha)$.\\

Comparing Eqs.~\eqref{eq:Xqu} and~\eqref{eq:Xclass}, we conclude that,
\be
{\cal F}(\alpha,t)
=
\lim_{\rm Th} \E[{\cal F}^{\rm qu}_\ell(\alpha,t)]
=
\lim_{\rm Th} {\cal F}_\ell^{\rm class}(\alpha, t),
\ee
and that ${\cal F}(\alpha,t)$ satisfies the nonlinear equation
\be \label{eq:F-round}
\partial_t {\cal F}
=
\frac{\alpha}{2}
-\alpha(1+\alpha)\,\partial_\alpha {\cal F}
+\frac{\alpha^2}{2}(1+\alpha)\big(\partial_\alpha {\cal F}\big)^2,
\ee
with initial condition ${\cal F}(\alpha,0)=\log(1+\alpha)$ and solution given by Eq.~\eqref{eq:fcs-quantum}.
\subsection{Finite-size corrections}
The previous result shows that, in the thermodynamic limit, the half-system charge full-counting statistics of the all-to-all QSSEP during domain-wall melting coincides with that of the corresponding classical process. This classical reducibility, already at finite times, is specific to transport observables and does not extend, for instance, to the thermodynamic entanglement entropy discussed in Sec.~\ref{sec:entanglement}, which has no classical counterpart. Finite-size corrections, however, distinguish the quantum and classical processes already at the level of transport, as encoded in the charge full-counting statistics.\\

To quantify such finite-size deviations, we introduce
\be
Q_L(\alpha,t):=\E[{\cal F}^{\rm qu}_\ell(\alpha,t)],
\qquad
C_L(\alpha,t):={\cal F}_\ell^{\rm class}(\alpha,t),
\ee
which, for $2\ell=L$, satisfy
\be
\partial_t Q_L = \mathscr{F}(\partial_\alpha Q_L) + r_L^{\rm qu},
\qquad
\partial_t C_L = \mathscr{F}(\partial_\alpha C_L) + r_L^{\rm class},
\ee
with $\mathscr{F}$ defined by the r.h.s. of \eqref{eq:F-round},
with common initial condition $Q_L(\alpha,0)=C_L(\alpha,0)=\log(1+\alpha)$, and remainders
\be
r_L^{\rm qu}={\cal O}(L^{-2}),
\qquad
r_L^{\rm class}={\cal O}(L^{-1}).
\ee

Defining $\Delta_L:=Q_L-C_L$, we obtain
\be
\partial_t \Delta_L
=
\mathscr{F}(\partial_\alpha Q_L)-\mathscr{F}(\partial_\alpha C_L)
+
(r_L^{\rm qu}-r_L^{\rm class}),
\qquad
\Delta_L(\alpha,0)=0.
\ee
Since $\partial_\alpha \Delta_L=\partial_\alpha Q_L-\partial_\alpha C_L$ and $\mathscr{F}$ is smooth, the nonlinear term can be written as
\be
\mathscr{F}(\partial_\alpha Q_L)-\mathscr{F}(\partial_\alpha C_L)
=
v_L(\alpha,t)\,\partial_\alpha \Delta_L,
\ee
where
\be
v_L(\alpha,t)
=
-\alpha(1+\alpha)
+\frac{\alpha^2}{2}(1+\alpha)\big(\partial_\alpha Q_L+\partial_\alpha C_L\big)
=
{\cal O}(1).
\ee
Hence,
\be
\partial_t \Delta_L
=
v_L(\alpha,t)\,\partial_\alpha \Delta_L
+
{\cal O}(L^{-1}),
\qquad
\Delta_L(\alpha,0)=0.
\ee

This is a transport equation with a forcing term of order $L^{-1}$. Since the initial condition is zero, transport alone does not generate deviations, and the difference is entirely controlled by the forcing. Therefore, for any fixed time,
\be
\sup_\alpha |\Delta_L(\alpha,t)| = {\cal O}(L^{-1}).
\ee

We conclude that
\be
\big|\E[{\cal F}_\ell^{\rm qu}(\alpha,t)] - {\cal F}_\ell^{\rm class}(\alpha,t)\big| = {\cal O}(L^{-1}).
\ee

\begin{figure}[t]
\centering
\includegraphics[width=\textwidth]{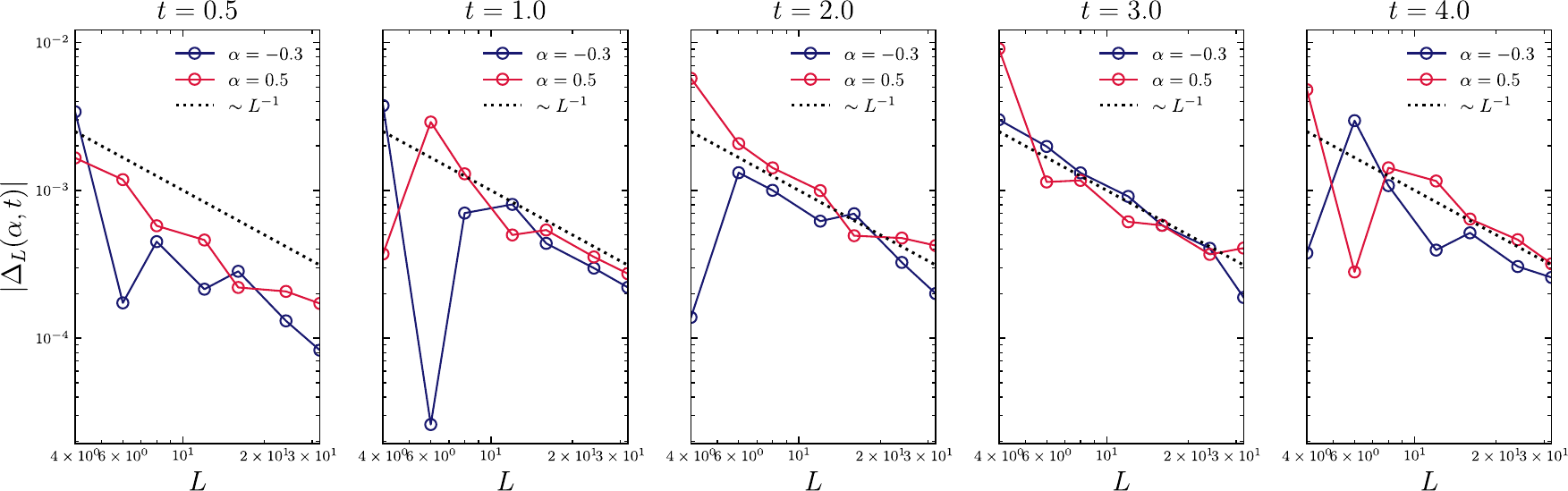}
\caption{Numerical analysis of the finite-size deviation $|\Delta_L(\alpha,t)|$ between the classical and quantum charge full-counting statistics as a function of $L$, for different values of $\alpha$ and $t$. Quantum data are obtained from simulations of the all-to-all QSSEP as described in Appendix~\ref{app:num}, while classical results are computed from the corresponding Markov process in Eq.~\eqref{eq:P-formal} via direct matrix exponentiation. For each system size, the number of samples is chosen as $N_{\rm samp}\sim{\cal O}(L^4)$, ensuring that statistical errors remain smaller than the finite-size deviations. The expected scaling $\sim L^{-1}$ is shown as a dotted black line for reference.}\label{fig:finite-size-corr}
\end{figure}

This scaling can be verified numerically, see Fig.~\ref{fig:finite-size-corr}. For the quantum process, $\E[{\cal F}_\ell(\alpha,t)]$ is estimated by averaging over $N_{\rm samp}$ realizations. Since the standard deviation scales as ${\cal O}(\sqrt{L})$, the statistical error behaves as ${\cal O}(\sqrt{L/N_{\rm samp}})$. Resolving a signal of order $L^{-1}$ therefore requires $N_{\rm samp}\gg L^{3}$, so that the statistical error remains much smaller than the finite-size correction of interest. In practice, this restricts the numerical analysis to modest system sizes.

\section{Outlook}\label{sec:outlook}
In this work we have studied  the melting of a domain wall in the QSSEP with all-to-all hoppings. By means of recent results in RMT, we have derived explicit and exact formulas for the bipartite entanglement entropy and the full counting statistics at all times. Different from previous approaches, we do not make use of replica methods but employ a mapping to the so-called Jacobi process, allowing us to obtain an exact description of all the moments of the eigenvalues of the subsystem correlation matrix. For the boundary-driven case, we mention a recent replica-based approach to the model in Eq.~\eqref{eq:dH}, extending steady-state results of the local model in Ref.~\cite{bernard2019open} to the all-to-all QSSEP \cite{Li2026}. Therefore, our work presents a new approach to stochastic dynamics, featuring a particularly nice and direct application of RMT constructions.

As we have highlighted, our analytic formulas allow for a rare exact comparison between particle fluctuations in the quantum and classical SSEP dynamics. Notably, we find that the quantum and classical particle full counting statistics coincide in the thermodynamic limit, with no finite time corrections. To our knowledge, such a comparison was previously only available in the QSSEP for late-time stationary states~\cite{bauer2024bernoulli,bernard2025large,albert2026universal,costa2025emergence}, with no finite-time result available. Therefore, our work motivates further investigations in this direction. It would be particularly interesting to understand whether finite-time corrections appear for different initial states or different connectivity of the underlying lattice, which could yield further insight into the role of coherent fluctuations on particle transport. We leave these questions to future work.

\section*{Acknowledgments}~ We acknowledge Andrea De Luca for useful discussions. This work was partially funded by the European Union (L.P.~--~ERC, QUANTHEM, 101114881; S.S.~--~MSCA, GENESYS, 101103348). Views and opinions expressed are however those of the author(s) only and do not necessarily reflect those of the European Union or the European Research Council Executive Agency. Neither the European Union nor the granting authority can be held responsible for them. We also acknowledge support from the CNRS, the ENS (D.B., S.S.), and the Simons Collaboration “Probabilistic Paths to QFT" (D.B.).

\medskip

%

\appendix
\renewcommand{\thesection}{\Alph{section}}
\section{Numerical simulations}\label{app:num}
We briefly describe the numerical procedure used to simulate the real-time dynamics of the all-to-all QSSEP defined in Sec.~\ref{sec:the_model}. We work directly at the level of the correlation matrix $\Gamma(t)$ introduced in Eq.~\eqref{eq:gamma_function}. For each realization, the initial condition is the domain-wall state~\eqref{eq:DW_state}, corresponding to
\be
\Gamma_{ij}(0)=\delta_{ij}\,\Theta(M-i),
\ee
where $\Theta(M-i)$ is the Heaviside theta function. The stochastic evolution is implemented by discretizing time, $t\to t+\Delta  t$. The the unitary increment corresponding to Eq.~\eqref{eq:gamma_function} is realized by setting
\be
U(\Delta t)\simeq \exp\big[-i\,\,\Delta t J(t)\big],
\ee
where
\be\label{eq:dw_mat_discrete}
\big(J(t)\big)_{\!\!jk}=\begin{cases}
\tfrac{1}{\sqrt{L}} (J_R^{jk}(t)+i J_I^{jk}(t)), \quad k\geq j;\\[12pt]
\tfrac{1}{\sqrt{L}} (J_R^{jk}(t)-i J_I^{jk}(t)), \quad k< j.
\end{cases}
\ee
Here $J_R^{jk}(t)$ and $J_I^{jk}(t)$ are real random variables drawn from a Gaussian distribution with vanishing mean and variance
\begin{equation}
    \mathbb{E}[J_R^{jk}(t)J_R^{jk}(t)]=\mathbb{E}[J_I^{jk}(t)J_I^{jk}(t)]=\frac{1}{\Delta t}\,.
\end{equation}

At each time step, a new realization of $J(t)$ is generated and the correlation matrix is then updated according to
\be
\Gamma(t+\Delta t)=U(\Delta t)\,\Gamma(t)\,U^\dagger(\Delta t),
\ee
which is the discrete-time counterpart of the stochastic evolution induced by Eq.~\eqref{eq:SDE_U}.\\

At each time step, we restrict $\Gamma(t)$ to the subsystem $A_\ell=\{1,\dots,\ell\}$, and compute its eigenvalues $\{\lambda_j(t)\}_{j=1}^\ell$. As discussed in Sec.~\ref{sec:gauss-evo}, these fully determine both the entanglement entropy and the charge full-counting statistics for each realization of the noise. The final results are then obtained by averaging over $N_{\rm samp}$ independent realizations of the noise,
\be
\overline{S_\ell(t)}
=
\frac{1}{N_{\rm samp}}\sum_{r=1}^{N_{\rm samp}} S_\ell^{(r)}(t),
\qquad
\overline{\log Z_\ell(\alpha,t)}
=
\frac{1}{N_{\rm samp}}\sum_{r=1}^{N_{\rm samp}} \log Z_\ell^{(r)}(\alpha,t).
\ee
The value of $\Delta t$ controls the accuracy of the numerical results. For all figures, we have verified that $\Delta t$ is small enough as not to cause discretization errors that are visible at the scales of the plots.
\section{Details on the calculations of the charge full-counting statistics}\label{app:algebra-fcs}
We provide additional details on the derivation of Eq.~\eqref{eq:fcs-quantum} for the case $M=\ell=L/2$. Using the expression of the moments in Eq.~\eqref{eq:explicit_moment}, the charge full-counting statistics can be rewritten as
\be
{\cal F}(\alpha,t)={\cal F}_{\rm ss}(\alpha)+\widetilde{\cal F}(\alpha,t),
\ee
with
\be
{\cal F}_{\rm ss}(\alpha)
:=
\sum_{n\ge1}\frac{(-1)^{n+1}}{n} \left(\frac{\alpha}{4}\right)^n \binom{2n}{n},
\ee
and
\be
\widetilde{\cal F}(\alpha,t)
:=
2\sum_{n\ge1}\frac{(-1)^{n+1}}{n}
\left(\frac{\alpha}{4}\right)^n
\sum_{k=1}^{n}\binom{2n}{n-k}\frac{1}{k}
L^{(1)}_{k-1}(2kt)\, e^{-kt}.
\ee

We first evaluate the stationary contribution. Using the identity
\be
\sum_{n\ge1}\frac{(-1)^{n+1}}{n}\binom{2n}{n}x^n
=
2\log\!\left(\frac{1+\sqrt{1+4x}}{2}\right),
\ee
one obtains
\be
{\cal F}_{\rm ss}(\alpha)
=
2\log\!\left(\frac{1+\sqrt{1+\alpha}}{2}\right).
\label{eq:Fss_from_central_binomial}
\ee

We now turn to the time-dependent contribution. Exchanging the order of summation over $k$ and $n$, one finds
\be
\widetilde{\cal F}(\alpha,t)
=
\sum_{k\ge1}\frac{2}{k}
L^{(1)}_{k-1}(2kt)\, e^{-kt}
\sum_{n\ge k}
\frac{(-1)^{n+1}}{n}
\binom{2n}{n-k}
\left(\frac{\alpha}{4}\right)^n.
\ee
Setting $n=m+k$, the inner sum becomes
\be
\sum_{m\ge0}\frac{(-1)^{m+k+1}}{m+k}
\binom{2m+2k}{m}
\left(\frac{\alpha}{4}\right)^{m+k}.
\ee
Using the identity
\be
\frac{1}{m+k}\binom{2m+2k}{m}
=
\frac{2^{2m}}{k}\frac{(2k)_m\left(k+\frac12\right)_m}{(2k+1)_m\,m!},
\ee
where $(k)_m$ denotes the Pochhammer symbol, together with
\be
\frac{(k)_m}{(k+1)_m}=\frac{k}{k+m},
\ee
and recalling the definition
\be
{}_2F_1(a,b;c;z):=\sum_{m\ge0} \frac{(a)_m (b)_m}{(c)_m}\,\frac{z^m}{m!},
\ee
one obtains
\be
\sum_{n\ge k}\frac{(-1)^{n+1}}{n}\binom{2n}{n-k}\left(\frac{\alpha}{4}\right)^n
=
-\left(-\frac{\alpha}{4}\right)^k
\,{}_2F_1\!\left(k+\frac12,k;2k+1;\alpha\right).
\ee
Therefore,
\be
\widetilde{\cal F}(\alpha,t)
=
-2\sum_{k\ge1}
e^{-kt}L^{(1)}_{k-1}(2kt)
\left(-\frac{\alpha}{4}\right)^k
\,{}_2F_1\!\left(k+\frac12,k;2k+1;\alpha\right).
\ee
Combining this with Eq.~\eqref{eq:Fss_from_central_binomial} yields Eq.~\eqref{eq:fcs-quantum}.
\section{Self averaging behavior}\label{app:self-avg}
In this appendix, we investigate the self-averaging of the following class of spectral observables
\be
X_f(t):=\frac{1}{\ell}\sum_{i=1}^\ell f(\lambda_i(t)),
\ee
where $\lambda_i(t)$ are the non-zero eigenvalues of $\Gamma_\ell(t)$, evolving according to the SDE in Eq.~\eqref{eq:lambda-jacobi}. Writing the SDE schematically as
\be
d\lambda_i(t)=A_i(t)\,dt+dB_i(t),
\ee
with drift 
\be
A_i(t):=\frac{1}{L}\left[\ell-L \lambda_{i}(t)+\sum_{j \neq i} \frac{\lambda_{i}(t)\left(1-\lambda_{j}(t)\right)+\lambda_{j}(t)\left(1-\lambda_{i}(t)\right)}{\lambda_{i}(t)-\lambda_{j}(t)}\right],
\ee
and noise
\be
dB_i(t):=\sqrt{\frac{2}{L}\lambda_i(1-\lambda_i)}\,d\nu_i(t),
\quad
(dB_i)^2=:\sigma_i(t)dt=\frac{2}{L}\lambda_i(1-\lambda_i)\,dt,
\ee
an application of It\=o calculus yields
\be
dX_f(t)=\tilde A_f(t)\,dt+d\tilde B_f(t),
\ee
with
\be
\tilde A_f(t)=\frac{1}{\ell}\sum_{i=1}^\ell\Big[f'(\lambda_i)A_i(t)+\tfrac12\sigma_i(t)f''(\lambda_i)\Big],
\quad
d\tilde B_f(t)=\frac{1}{\ell}\sum_{i=1}^\ell f'(\lambda_i)dB_i(t),
\ee
and quadratic variation
\be
(d\tilde B_f)^2=\tilde\sigma_f(t)\,dt,
\quad
\tilde\sigma_f(t)=\frac{1}{\ell^2}\sum_{i=1}^\ell (f'(\lambda_i))^2\,\frac{2}{L}\lambda_i(1-\lambda_i).
\ee

Assuming $f\in C^2([0,1])$ with bounded derivatives, one has
\be
\tilde\sigma_f(t)\le \frac{\|f'\|_\infty^2}{2L\ell}.
\ee
Using It\=o calculus for $X_f^2$ and taking expectations, one finds
\be
\frac{d}{dt}\Var(X_f(t))
=
2\Big(\E[X_f\tilde A_f]-\E[X_f]\E[\tilde A_f]\Big)
+\E[\tilde\sigma_f].
\ee
Since $\tilde A_f$ is a smooth bounded function of the eigenvalues $\lambda_i(t)$, one has
\be
\big|\E[X_f\tilde A_f]-\E[X_f]\E[\tilde A_f]\big|
\le C_f\,\Var(X_f),
\ee
for some constant $C_f$. It follows that
\be
\frac{d}{dt}\Var(X_f(t))
\le
2C_f\,\Var(X_f(t))
+\frac{\|f'\|_\infty^2}{2L\ell}.
\ee
With deterministic initial condition, Gr\"onwall's inequality yields
\be
\Var(X_f(t))={\cal O}\!\left(\frac{1}{L\ell}\right).
\ee
In particular, for $\ell\sim L$,
\be
\Var(X_f(t))={\cal O}(L^{-2}),
\ee
so that $X_f(t)$ is self-averaging.\\

The observables of interest fall within this class. In particular, the R\'enyi entropies and the charge full-counting statistics correspond to
\be
f_q(\lambda)=\frac{1}{1-q}\log\!\big(\lambda^q+(1-\lambda)^q\big),
\qquad
f_\alpha(\lambda)=\log(1+\alpha\lambda),
\ee
which have bounded derivatives on $[0,1]$, so that the above result applies directly. More generally, the argument requires that
\be
\lambda(1-\lambda)\big(f'(\lambda)\big)^2,
\qquad
\lambda(1-\lambda)|f''(\lambda)|
\ee
remain bounded. This condition is also satisfied by the von Neumann entropy,
\be
f_{\rm vN}(\lambda)=-\lambda\log\lambda-(1-\lambda)\log(1-\lambda),
\ee
for which the divergences of $f'$ and $f''$ at the endpoints are compensated by the factor $\lambda(1-\lambda)$. Hence the entanglement entropy is self-averaging as well.\\

\begin{figure}[t]
\centering
\includegraphics[width=0.49\textwidth]{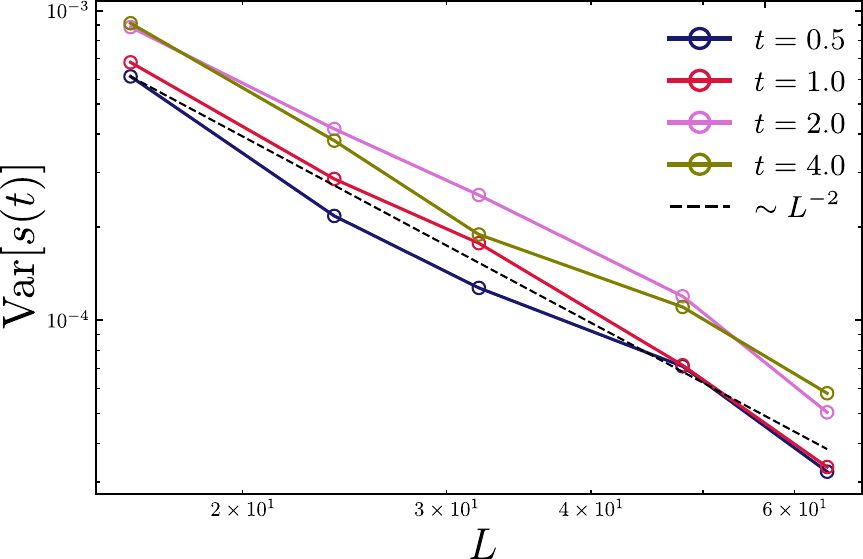}\hspace{.15cm}\includegraphics[width=0.5\textwidth]{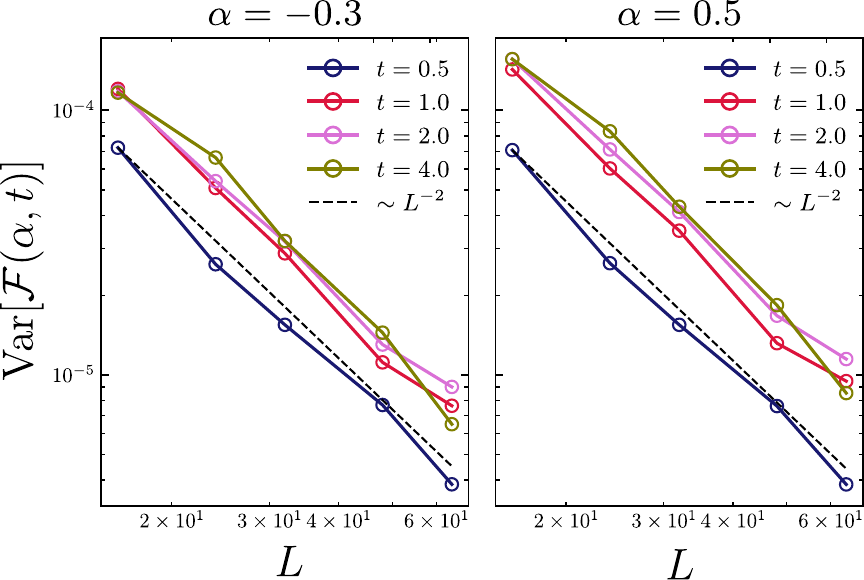}
\caption{Sample-to-sample variance \eqref{eq:sample-var} for the entanglement (left panel) and charge full-counting statistics (right panels) as a function of $L$. The dashed line indicates the expected scaling $\sim L^{-2}$. In these figures, we set $N_{\rm samp}=200$.}\label{fig:self-avg2}
\end{figure}

Self-averaging can be tested numerically by estimating the sample average over $N_{\rm samp}$ independent realizations,
\be\label{eq:sample-avg}
\overline{X_f(t)}=\frac{1}{N_{\rm samp}}\sum_{r=1}^{N_{\rm samp}} X_f^{(r)}(t).
\ee
A quantitative measure is provided by the sample-to-sample variance
\be\label{eq:sample-var}
{\rm Var}[X_f]=
\frac{1}{N_{\rm samp}}
\sum_{r=1}^{N_{\rm samp}}
\left(X_f^{(r)}-\overline{X_f}\right)^2.
\ee
Self-averaging implies ${\rm Var}[X_f]\to 0$ as $L\to\infty$, with the scaling ${\rm Var}[X_f]\sim L^{-2}$ predicted by the analytical argument. This behavior is confirmed numerically in Fig.~\ref{fig:self-avg2} for both the entanglement entropy and the full counting statistics.
\section{Finite-time calculation of the lowest cumulants}\label{app:low-cum-ssep}
In this appendix, we provide a direct calculation of the mean and variance of the subsystem occupation number $N_\ell(t)$ in the all-to-all SSEP with $L$ sites and $M$ particles. The dynamics of $N_\ell$ is a birth–death process with rates given in Eq.~\eqref{eq:birthdeath_rates}. Accordingly, the mean satisfies
\be
\frac{d}{dt}\E[N_\ell(t)]
=
\E[b_{N_\ell}-d_{N_\ell}]
=
\frac{\ell M}{L}-\E[N_\ell(t)],
\ee
which yields
\be\label{eq:classical_mean_general_app}
\E[N_\ell(t)]
=
\frac{\ell M}{L}
+
\left(\E[N_\ell(0)]-\frac{\ell M}{L}\right)e^{-t}.
\ee
For the domain-wall initial condition with $M=\ell=L/2$, this reduces to
\be\label{eq:classical_mean_app}
\E[N_\ell(t)]
=
\frac{\ell}{2}(1+e^{-t}).
\ee

The second moment satisfies a closed equation, from which one obtains the variance
\be
V^{\rm class}_\ell(t):=\Var(N_\ell(t)).
\ee
In the case $M=\ell=L/2$, one finds
\be
\frac{d}{dt}V^{\rm class}_\ell(t)
=
-2\left(1-\frac{1}{L}\right)V^{\rm class}_\ell(t)
+
\frac{L}{8}(1+e^{-2t}),
\ee
with initial condition $V^{\rm class}_\ell(0)=0$. Solving this equation and taking the thermodynamic limit yields
\be\label{eq:classical_variance_app}
V^{\rm class}_\ell(t)=\frac{L}{16}\left[1+(2t-1)e^{-2t}\right]\left(1+{\cal O}(L^{-1})\right).
\ee

We now extract the first two cumulants from the quantum generating function associated with the all-to-all QSSEP. Using Eq.~\eqref{eq:fcs-quantum-moments},
\be
{\cal F}(\alpha,t)
=
\sum_{n\ge1}\frac{(-1)^{n+1}}{n}\alpha^n m_n(t),
\ee
the cumulants can be expressed as
\be
\frac{\kappa_1^{\rm qu}(t)}{M}=m_1(t),
\qquad
\frac{\kappa_2^{\rm qu}(t)}{M}=m_1(t)-m_2(t).
\ee
For $M=\ell=L/2$, one has
\be
m_1(t)=\frac12(1+e^{-t}),
\qquad
m_2(t)=\frac{3}{8}+\frac12 e^{-t}+\left(\frac18-\frac{t}{4}\right)e^{-2t}.
\ee
This yields
\be\label{eq:quantum_mean_app}
\kappa_1^{\rm qu}(t)
=
\E[N_\ell(t)]
=
\frac{\ell}{2}(1+e^{-t}),
\ee
and
\be\label{eq:quantum_variance_app}
\kappa_2^{\rm qu}(t)
=
\frac{L}{16}\left[1+(2t-1)e^{-2t}\right].
\ee

We conclude that, in the thermodynamic limit, the mean and variance of the classical and quantum processes coincide exactly, already at finite times.
\section{Details of the derivation of Eqs.~\eqref{eq:tmp1} and \eqref{eq:tmp2}}\label{app:algebra2}
In this appendix, we provide the details of the derivation of Eqs.~\eqref{eq:tmp1} and \eqref{eq:tmp2}, which govern the exact evolution of the charge full-counting statistics in the all-to-all QSSEP. We begin by recalling the definition of the finite-size charge full-counting statistics,
\be
X_\alpha(t)\equiv {\cal F}^{\rm qu}_\ell(\alpha,t)
=\frac{1}{\ell}\sum_{i=1}^{\ell}\log\big(1+\alpha\lambda_i(t)\big),
\ee
which we denote by $X_\alpha(t)$ for brevity in this section. Using It\=o calculus together with the SDE $d\lambda_i(t)= A_i(t)\ dt + dB_i(t)$ (cf. Eq.~\eqref{eq:lambda-jacobi} and Appendix~\ref{app:self-avg}), one finds
\be
dX_\alpha
=
\frac{1}{\ell}\sum_{i=1}^{\ell}
\frac{\alpha}{1+\alpha\lambda_i}\,d\lambda_i
-\frac{1}{2\ell}\sum_{i=1}^{\ell}
\frac{\alpha^2}{(1+\alpha\lambda_i)^2}\,\sigma_i\,dt,
\ee
with $\sigma=2\lambda_i(1-\lambda_i)/L$ as in Appendix~\ref{app:self-avg}. This equation can be schematically written as
\be
dX_\alpha=\tilde A_\alpha\,dt+d\tilde B_\alpha.
\ee
Here,
\be
\tilde A_\alpha
:=
\frac{1}{\ell}\sum_{i=1}^{\ell}
\frac{\alpha}{1+\alpha\lambda_i}A_i
-\frac{1}{2\ell}\sum_{i=1}^{\ell}
\frac{\alpha^2}{(1+\alpha\lambda_i)^2}\sigma_i,
\quad
d\tilde B_\alpha:=
\frac{1}{\ell}\sum_{i=1}^{\ell}
\frac{\alpha}{1+\alpha\lambda_i}dB_i.
\ee
Taking expectation yields Eq.~\eqref{eq:tmp1},
\be
\de_t\E[X_\alpha(t)]=\E[\tilde A_\alpha(t)].
\ee

We now analyze the drift term. Using the explicit expressions of $A_i$ and $\sigma_i$, one obtains
\begin{align}
\tilde A_\alpha(t)
&=\frac{1}{\ell L}\sum_{i=1}^{\ell}
\left[
\frac{\alpha(\ell-L\lambda_i)}{1+\alpha\lambda_i}
-
\frac{\alpha^2}{(1+\alpha\lambda_i)^2}\lambda_i(1-\lambda_i)
\right]\nonumber\\[4pt]
&\quad
+\frac{1}{\ell L}\sum_{i=1}^{\ell}
\frac{\alpha}{1+\alpha\lambda_i}
\sum_{j\neq i}
\frac{\lambda_i(1-\lambda_j)+\lambda_j(1-\lambda_i)}{\lambda_i-\lambda_j}.
\end{align}

The double sum can be symmetrized by pairing $(i,j)$ and $(j,i)$. Defining $g_\alpha(\lambda)=\alpha/(1+\alpha\lambda)$, one finds
\be\label{eq:aux}
\sum_{i=1}^{\ell} g_\alpha(\lambda_i)\sum_{j\neq i}
\frac{\lambda_i(1-\lambda_j)+\lambda_j(1-\lambda_i)}{\lambda_i-\lambda_j}
=
-\alpha^2\sum_{i<j}
\frac{\lambda_i(1-\lambda_j)+\lambda_j(1-\lambda_i)}
{(1+\alpha\lambda_i)(1+\alpha\lambda_j)}.
\ee
Rewriting the pair sum in terms of products of single sums yields
\be
\eqref{eq:aux}=
-\alpha^2
\left(\sum_{i=1}^{\ell}\frac{\lambda_i}{1+\alpha\lambda_i}\right)
\left(\sum_{j=1}^{\ell}\frac{1-\lambda_j}{1+\alpha\lambda_j}\right)
+\alpha^2\sum_{i=1}^{\ell}\frac{\lambda_i(1-\lambda_i)}{(1+\alpha\lambda_i)^2}.
\ee
The last term cancels the explicit It\=o correction, and the drift simplifies to
\be
\tilde A_\alpha(t)
=
\frac{1}{\ell L}\sum_{i=1}^{\ell}
\frac{\alpha(\ell-L\lambda_i)}{1+\alpha\lambda_i}
-\frac{\alpha^2}{\ell L}
\left(\sum_{i=1}^{\ell}\frac{\lambda_i}{1+\alpha\lambda_i}\right)
\left(\sum_{j=1}^{\ell}\frac{1-\lambda_j}{1+\alpha\lambda_j}\right).
\ee

We thus obtain the exact evolution equation in \eqref{eq:tmp2}
\be
\de_t\E[X_\alpha(t)]
=
\frac{1}{\ell L}\E\!\left[\sum_{i=1}^{\ell}
\frac{\alpha(\ell-L\lambda_i)}{1+\alpha\lambda_i}\right]
-\frac{\alpha^2}{\ell L}
\E\!\left[
\left(\sum_{i=1}^{\ell}\frac{\lambda_i}{1+\alpha\lambda_i}\right)
\left(\sum_{j=1}^{\ell}\frac{1-\lambda_j}{1+\alpha\lambda_j}\right)
\right].
\ee
\bigskip
\hrulefill
\bibliographystyle{iopart-num}
\bibliography{bibliography}

\providecommand{\newblock}{}
\begin{thebibliography}{10}
\expandafter\ifx\csname url\endcsname\relax
  \def\url#1{{\tt #1}}\fi
\expandafter\ifx\csname urlprefix\endcsname\relax\def\urlprefix{URL }\fi
\providecommand{\eprint}[2][]{\url{#2}}

\bibitem{breuer2002theory}
Breuer H~P and Petruccione F 2002 {\em The theory of open quantum systems\/}
  (Oxford University Press on Demand)

\bibitem{nahum2017quantum}
Nahum A, Ruhman J, Vijay S and Haah J 2017 {\em Phys. Rev. X\/} {\bf 7}(3)
  031016 \urlprefix\url{https://link.aps.org/doi/10.1103/PhysRevX.7.031016}

\bibitem{potter_entanglement_2022}
Potter A~C and Vasseur R 2022 Entanglement {Dynamics} in {Hybrid} {Quantum}
  {Circuits} {\em Entanglement in {Spin} {Chains}: {From} {Theory} to {Quantum}
  {Technology} {Applications}\/} ed Bayat A, Bose S and Johannesson H (Cham:
  Springer International Publishing) pp 211--249 ISBN 978-3-031-03998-0
  \urlprefix\url{https://doi.org/10.1007/978-3-031-03998-0_9}

\bibitem{fisher2023random}
Fisher M~P, Khemani V, Nahum A and Vijay S 2023 {\em Ann. Rev. Cond. Matt.
  Phys.\/} {\bf 14} 335--379 ISSN 1947-5462
  \urlprefix\url{https://www.annualreviews.org/content/journals/10.1146/annurev-conmatphys-031720-030658}

\bibitem{rakovszky2019sub}
Rakovszky T, Pollmann F and von Keyserlingk C~W 2019 {\em Phys. Rev. Lett.\/}
  {\bf 122}(25) 250602
  \urlprefix\url{https://link.aps.org/doi/10.1103/PhysRevLett.122.250602}

\bibitem{emergent2019zhou}
Zhou T and Nahum A 2019 {\em Phys. Rev. B\/} {\bf 99}(17) 174205
  \urlprefix\url{https://link.aps.org/doi/10.1103/PhysRevB.99.174205}

\bibitem{gullans2019entanglement}
Gullans M~J and Huse D~A 2019 {\em Phys. Rev. X\/} {\bf 9}(2) 021007
  \urlprefix\url{https://link.aps.org/doi/10.1103/PhysRevX.9.021007}

\bibitem{bertini2019entanglement}
Bertini B, Kos P and Prosen T 2019 {\em Phys. Rev. X\/} {\bf 9}(2) 021033
  \urlprefix\url{https://link.aps.org/doi/10.1103/PhysRevX.9.021033}

\bibitem{znidarivc2020entanglement}
Znidaric M 2020 {\em Comm. Phys.\/} {\bf 3} 1--9
  \urlprefix\url{https://www.nature.com/articles/s42005-020-0366-7}

\bibitem{huang2020dynamics}
Huang Y 2020 {\em IOP SciNotes\/} {\bf 1} 035205
  \urlprefix\url{https://iopscience.iop.org/article/10.1088/2633-1357/abd1e2}

\bibitem{entanglement2020zhou}
Zhou T and Nahum A 2020 {\em Phys. Rev. X\/} {\bf 10}(3) 031066
  \urlprefix\url{https://link.aps.org/doi/10.1103/PhysRevX.10.031066}

\bibitem{nahum2018operator}
Nahum A, Vijay S and Haah J 2018 {\em Phys. Rev. X\/} {\bf 8}(2) 021014
  \urlprefix\url{https://link.aps.org/doi/10.1103/PhysRevX.8.021014}

\bibitem{vonKeyserlingk2018operator}
von Keyserlingk C~W, Rakovszky T, Pollmann F and Sondhi S~L 2018 {\em Phys.
  Rev. X\/} {\bf 8}(2) 021013
  \urlprefix\url{https://link.aps.org/doi/10.1103/PhysRevX.8.021013}

\bibitem{chan2018Solution}
Chan A, De~Luca A and Chalker J~T 2018 {\em Phys. Rev. X\/} {\bf 8}(4) 041019
  \urlprefix\url{https://link.aps.org/doi/10.1103/PhysRevX.8.041019}

\bibitem{sunderhauf2018localization}
S\"underhauf C, P\'erez-Garc\'ia D, Huse D~A, Schuch N and Cirac J~I 2018 {\em
  Phys. Rev. B\/} {\bf 98}(13) 134204
  \urlprefix\url{https://link.aps.org/doi/10.1103/PhysRevB.98.134204}

\bibitem{diffusive2018rakovszky}
Rakovszky T, Pollmann F and von Keyserlingk C~W 2018 {\em Phys. Rev. X\/} {\bf
  8}(3) 031058
  \urlprefix\url{https://link.aps.org/doi/10.1103/PhysRevX.8.031058}

\bibitem{khemani2018operator}
Khemani V, Vishwanath A and Huse D~A 2018 {\em Phys. Rev. X\/} {\bf 8}(3)
  031057 \urlprefix\url{https://link.aps.org/doi/10.1103/PhysRevX.8.031057}

\bibitem{hunter2018operator}
Hunter-Jones N 2018 {\em arXiv:1812.08219\/}
  \urlprefix\url{https://arxiv.org/abs/1812.08219}

\bibitem{friedman2019spectral}
Friedman A~J, Chan A, De~Luca A and Chalker J~T 2019 {\em Phys. Rev. Lett.\/}
  {\bf 123}(21) 210603
  \urlprefix\url{https://link.aps.org/doi/10.1103/PhysRevLett.123.210603}

\bibitem{hosur2016chaos}
Hosur P, Qi X~L, Roberts D~A and Yoshida B 2016 {\em JHEP\/} {\bf 2016} 4
  \urlprefix\url{https://link.springer.com/article/10.1007%2FJHEP02%282016%29004}

\bibitem{scrambling2020bertini}
Bertini B and Piroli L 2020 {\em Phys. Rev. B\/} {\bf 102}(6) 064305
  \urlprefix\url{https://link.aps.org/doi/10.1103/PhysRevB.102.064305}

\bibitem{piroli2020random}
Piroli L, S\"underhauf C and Qi X~L 2020 {\em JHEP\/} {\bf 2020} 1--35
  \urlprefix\url{https://link.springer.com/article/10.1007/JHEP04(2020)063}

\bibitem{bauer2017stochastic}
Bauer M, Bernard D and Jin T 2017 {\em SciPost Phys.\/} {\bf 3}(5) 033
  \urlprefix\url{https://scipost.org/10.21468/SciPostPhys.3.5.033}

\bibitem{onorati2017mixing}
Onorati E, Buerschaper O, Kliesch M, Brown W, Werner A and Eisert J 2017 {\em
  Comm. Math. Phys.\/} {\bf 355} 905--947
  \urlprefix\url{https://link.springer.com/article/10.1007%2Fs00220-017-2950-6}

\bibitem{knap2018entanglement}
Knap M 2018 {\em Phys. Rev. B\/} {\bf 98}(18) 184416
  \urlprefix\url{https://link.aps.org/doi/10.1103/PhysRevB.98.184416}

\bibitem{rowlands2018noisy}
Rowlands D~A and Lamacraft A 2018 {\em Phys. Rev. B\/} {\bf 98}(19) 195125
  \urlprefix\url{https://link.aps.org/doi/10.1103/PhysRevB.98.195125}

\bibitem{zhou2019operator}
Zhou T and Chen X 2019 {\em Phys. Rev. E\/} {\bf 99}(5) 052212
  \urlprefix\url{https://link.aps.org/doi/10.1103/PhysRevE.99.052212}

\bibitem{sunderhauf2019quantum}
S\"underhauf C, Piroli L, Qi X~L, Schuch N and Cirac J~I 2019 {\em JHEP\/} {\bf
  2019} 1--44
  \urlprefix\url{https://link.springer.com/article/10.1007%2FJHEP11%282019%29038}

\bibitem{bernard2020entanglement}
Bernard D and Le~Doussal P 2020 {\em EPL\/} {\bf 131} 10007
  \urlprefix\url{https://iopscience.iop.org/article/10.1209/0295-5075/131/10007}

\bibitem{bauer2019equilibrium}
Bauer M, Bernard D and Jin T 2019 {\em SciPost Phys.\/} {\bf 6}(4) 45
  \urlprefix\url{https://scipost.org/10.21468/SciPostPhys.6.4.045}

\bibitem{barraquand2025introduction}
Barraquand G and Bernard D 2025 {\em arXiv:2507.01570\/}
  \urlprefix\url{https://arxiv.org/abs/2507.01570}

\bibitem{kipnis1989hydrodynamics}
Kipnis C, Olla S and Varadhan S 1989 {\em Comm. Pure App. Math.\/} {\bf 42}
  115--137
  \urlprefix\url{https://onlinelibrary.wiley.com/doi/abs/10.1002/cpa.3160420202}

\bibitem{spohn2012large}
Spohn H 2012 {\em Large scale dynamics of interacting particles\/} (Springer
  Science \& Business Media)

\bibitem{derrida1993exact}
Derrida B, Evans M~R, Hakim V and Pasquier V 1993 {\em J. Phys. A: Math.
  Gen.\/} {\bf 26} 1493--1517
  \urlprefix\url{https://iopscience.iop.org/article/10.1088/0305-4470/26/7/011}

\bibitem{derrida2007non}
Derrida B 2007 {\em J. Stat. Mech.\/} {\bf 2007} P07023
  \urlprefix\url{https://iopscience.iop.org/article/10.1088/1742-5468/2007/07/P07023}

\bibitem{bodineau2004current}
Bodineau T and Derrida B 2004 {\em Phys. Rev. Lett.\/} {\bf 92}(18) 180601
  \urlprefix\url{https://link.aps.org/doi/10.1103/PhysRevLett.92.180601}

\bibitem{mallick2015exclusion}
Mallick K 2015 {\em Physica A: Stat. Mech. Appl.\/} {\bf 418} 17--48
  \urlprefix\url{https://www.sciencedirect.com/science/article/abs/pii/S0378437114006189?via%3Dihub}

\bibitem{bravyi2004lagrangian}
Bravyi S 2004 {\em arXiv quant-ph/0404180\/}
  \urlprefix\url{https://arxiv.org/abs/quant-ph/0404180}

\bibitem{surace2022tagliacozzo}
Surace J and Tagliacozzo L 2022 {\em SciPost Phys. Lect. Notes\/}  54
  \urlprefix\url{https://scipost.org/10.21468/SciPostPhysLectNotes.54}

\bibitem{bernard2019open}
Bernard D and Jin T 2019 {\em Phys. Rev. Lett.\/} {\bf 123}(8) 080601
  \urlprefix\url{https://link.aps.org/doi/10.1103/PhysRevLett.123.080601}

\bibitem{jin2020stochastic}
Jin T, Krajenbrink A and Bernard D 2020 {\em Phys. Rev. Lett.\/} {\bf 125}(4)
  040603
  \urlprefix\url{https://link.aps.org/doi/10.1103/PhysRevLett.125.040603}

\bibitem{alba2025nu}
Alba V 2025 {\em arXiv:2507.11674\/}
  \urlprefix\url{https://arxiv.org/abs/2507.11674}

\bibitem{bernard2025large}
Bernard D, Jin T, Scopa S and Wei S 2025 {\em Phys. Rev. E\/} {\bf 112}(3)
  034106 \urlprefix\url{https://link.aps.org/doi/10.1103/fs7f-z3k6}

\bibitem{bertini2001fluctuations}
Bertini L, De~Sole A, Gabrielli D, Jona-Lasinio G and Landim C 2001 {\em Phys.
  Rev. Lett.\/} {\bf 87}(4) 040601
  \urlprefix\url{https://link.aps.org/doi/10.1103/PhysRevLett.87.040601}

\bibitem{bertini2002macroscopic}
Bertini L, De~Sole A, Gabrielli D, Jona-Lasinio G and Landim C 2002 {\em J.
  Stat. Phys.\/} {\bf 107} 635--675
  \urlprefix\url{https://iopscience.iop.org/article/10.1088/1742-5468/2007/07/P07023}

\bibitem{bertini2015macroscopic}
Bertini L, De~Sole A, Gabrielli D, Jona-Lasinio G and Landim C 2015 {\em Rev.
  Mod. Phys.\/} {\bf 87}(2) 593--636
  \urlprefix\url{https://link.aps.org/doi/10.1103/RevModPhys.87.593}

\bibitem{bernard2021solution}
Bernard D and Jin T 2021 {\em Communications in Mathematical Physics\/} {\bf
  384} 1141--1185 ISSN 1432-0916
  \urlprefix\url{https://doi.org/10.1007/s00220-021-04087-x}

\bibitem{bernard2021entanglement}
Bernard D and Piroli L 2021 {\em Phys. Rev. E\/} {\bf 104}(1) 014146
  \urlprefix\url{https://link.aps.org/doi/10.1103/PhysRevE.104.014146}

\bibitem{russotto2026dynamics}
Russotto A, Ares F, Calabrese P and Alba V 2026 {\em J. Stat. Mech.\/} {\bf
  2026} 033103
  \urlprefix\url{https://iopscience.iop.org/article/10.1088/1742-5468/ae4bb9/meta}

\bibitem{russotto2026inhomogeneous}
Russotto A, Ares F, Calabrese P and Alba V 2026 {\em arXiv:2602.15122\/}
  \urlprefix\url{https://arxiv.org/abs/2602.15122}

\bibitem{hruza2023coherent}
Hruza L and Bernard D 2023 {\em Phys. Rev. X\/} {\bf 13}(1) 011045
  \urlprefix\url{https://link.aps.org/doi/10.1103/PhysRevX.13.011045}

\bibitem{bernard2023exact}
Bernard D and Hruza L 2023 {\em SciPost Phys.\/} {\bf 15} 175
  \urlprefix\url{https://scipost.org/10.21468/SciPostPhys.15.4.175}

\bibitem{bauer2024bernoulli}
Bauer M, Bernard D, Biane P and Hruza L 2024 Bernoulli variables, classical
  exclusion processes and free probability {\em Ann, Henri Poincar{\'e}\/}
  vol~25 (Springer) pp 125--172
  \urlprefix\url{https://link.springer.com/article/10.1007/s00023-023-01320-2}

\bibitem{albert2026universal}
Albert M, Bernard D, Jin T, Scopa S and Wei S 2026 {\em arXiv:2601.16883\/}
  \urlprefix\url{https://arxiv.org/abs/2601.16883}

\bibitem{bernard2022dynamics}
Bernard D, Essler F~H~L, Hruza L and Medenjak M 2022 {\em SciPost Phys.\/} {\bf
  12} 042 \urlprefix\url{https://scipost.org/10.21468/SciPostPhys.12.1.042}

\bibitem{swann2025spacetime}
Swann T, Bernard D and Nahum A 2025 {\em Phys. Rev. B\/} {\bf 112}(6) 064301
  \urlprefix\url{https://link.aps.org/doi/10.1103/PhysRevB.112.064301}

\bibitem{sachdev1993gapless}
Sachdev S and Ye J 1993 {\em Phys. Rev. Lett.\/} {\bf 70}(21) 3339--3342
  \urlprefix\url{https://link.aps.org/doi/10.1103/PhysRevLett.70.3339}

\bibitem{maldacena2016remarks}
Maldacena J and Stanford D 2016 {\em Phys. Rev. D\/} {\bf 94}(10) 106002
  \urlprefix\url{https://link.aps.org/doi/10.1103/PhysRevD.94.106002}

\bibitem{Fukai2026}
Fukai K and Katsura H 2026 {\em Physical Review B\/} {\bf 113} ISSN 2469-9969
  \urlprefix\url{http://dx.doi.org/10.1103/g4b2-n9c8}

\bibitem{Antal1999}
Antal T, Rácz Z, Rákos A and Sch\"{u}tz G~M 1999 {\em Physical Review E\/}
  {\bf 59} 4912–4918 ISSN 1095-3787
  \urlprefix\url{http://dx.doi.org/10.1103/PhysRevE.59.4912}

\bibitem{Antal2008}
Antal T, Krapivsky P~L and Rákos A 2008 {\em Physical Review E\/} {\bf 78}
  ISSN 1550-2376 \urlprefix\url{http://dx.doi.org/10.1103/PhysRevE.78.061115}

\bibitem{Karevski2002}
Karevski D 2002 {\em The European Physical Journal B - Condensed Matter\/} {\bf
  27} 147–152 ISSN 1434-6036
  \urlprefix\url{http://dx.doi.org/10.1140/epjb/e20020139}

\bibitem{Platini2007}
Platini T and Karevski D 2007 {\em Journal of Physics A: Mathematical and
  Theoretical\/} {\bf 40} 1711–1726 ISSN 1751-8121
  \urlprefix\url{http://dx.doi.org/10.1088/1751-8113/40/8/002}

\bibitem{Hunyadi2004}
Hunyadi V, Rácz Z and Sasvári L 2004 {\em Physical Review E\/} {\bf 69} ISSN
  1550-2376 \urlprefix\url{http://dx.doi.org/10.1103/PhysRevE.69.066103}

\bibitem{Collura2018}
Collura M, De~Luca A and Viti J 2018 {\em Physical Review B\/} {\bf 97} ISSN
  2469-9969 \urlprefix\url{http://dx.doi.org/10.1103/PhysRevB.97.081111}

\bibitem{Gruber2019}
Gruber M and Eisler V 2019 {\em Physical Review B\/} {\bf 99} ISSN 2469-9969
  \urlprefix\url{http://dx.doi.org/10.1103/PhysRevB.99.174403}

\bibitem{Eisler2018}
Eisler V and Maislinger F 2018 {\em Physical Review B\/} {\bf 98} ISSN
  2469-9969 \urlprefix\url{http://dx.doi.org/10.1103/PhysRevB.98.161117}

\bibitem{Scopa2021}
Scopa S, Krajenbrink A, Calabrese P and Dubail J 2021 {\em Journal of Physics
  A: Mathematical and Theoretical\/} {\bf 54} 404002 ISSN 1751-8121
  \urlprefix\url{http://dx.doi.org/10.1088/1751-8121/ac20ee}

\bibitem{Ares2022}
Ares F, Scopa S and Wald S 2022 {\em Journal of Physics A: Mathematical and
  Theoretical\/} {\bf 55} 375301 ISSN 1751-8121
  \urlprefix\url{http://dx.doi.org/10.1088/1751-8121/ac8209}

\bibitem{capizzi2023domain}
Capizzi L, Scopa S, Rottoli F and Calabrese P 2023 {\em Europhysics Letters\/}
  {\bf 141} 31002
  \urlprefix\url{https://iopscience.iop.org/article/10.1209/0295-5075/acb50a}

\bibitem{Scopa2023}
Scopa S and Karevski D 2023 {\em The European Physical Journal Special
  Topics\/} {\bf 232} 1763–1781 ISSN 1951-6401
  \urlprefix\url{http://dx.doi.org/10.1140/epjs/s11734-023-00845-1}

\bibitem{tiutiakina2025field}
Tiutiakina A, L{\'o}io H, Giachetti G, De~Nardis J and De~Luca A 2025 {\em
  Quantum\/} {\bf 9} 1794
  \urlprefix\url{https://quantum-journal.org/papers/q-2025-07-14-1794/}

\bibitem{oksendal2003stochastic}
Oksendal B 2003 {\em Stochastic Differential Equations: an Introduction with
  Applications\/} (Springer-Verlag Berlin Heidelberg)

\bibitem{surace2022fermionic}
Surace J and Tagliacozzo L 2022 {\em SciPost Phys. Lect. Notes\/}  54
  \urlprefix\url{https://scipost.org/10.21468/SciPostPhysLectNotes.54}

\bibitem{vidal2003entanglement}
Vidal G, Latorre J~I, Rico E and Kitaev A 2003 {\em Phys. Rev. Lett.\/} {\bf
  90}(22) 227902
  \urlprefix\url{https://link.aps.org/doi/10.1103/PhysRevLett.90.227902}

\bibitem{doumerc2005matrices}
Doumerc Y 2005 {\em Matrices al{\'e}atoires, processus stochastiques et groupes
  de r{\'e}flexions\/} Ph.D. thesis

\bibitem{deleaval2018moments}
Deleaval L and Demni N 2018 {\em J. Theor. Prob.\/} {\bf 31} 1759--1778
  \urlprefix\url{https://link.springer.com/article/10.1007/s10959-017-0761-5}

\bibitem{demni2008free}
Demni N 2008 {\em J. Theor. Prob.\/} {\bf 21} 118--143
  \urlprefix\url{https://link.springer.com/article/10.1007/s10959-007-0110-1}

\bibitem{collins2018spectral}
Collins B, Dahlqvist A and Kemp T 2018 {\em Prob. Theory Rel. Fields\/} {\bf
  170} 49--93
  \urlprefix\url{https://link.springer.com/article/10.1007/s00440-016-0753-x}

\bibitem{demni2018inverse}
Demni N and Hamdi T 2018 {\em Random Mat.: Th. App.\/} {\bf 7} 1850001
  \urlprefix\url{https://www.worldscientific.com/doi/abs/10.1142/S2010326318500016}

\bibitem{demni2020hermitian}
Demni N, Hamdi T and Souissi A 2020 {\em Funct. An. Appl.\/} {\bf 54} 257--271
  \urlprefix\url{https://link.springer.com/article/10.1134/S0016266320040036}

\bibitem{demni2012spectral}
Demni N, Hamdi T and Hmidi T 2012 {\em Ind. Univ. Math. J\/}  1351--1368
  \urlprefix\url{https://www.jstor.org/stable/24904084}

\bibitem{liu2018quantum}
Liu C, Chen X and Balents L 2018 {\em Phys. Rev. B\/} {\bf 97}(24) 245126
  \urlprefix\url{https://link.aps.org/doi/10.1103/PhysRevB.97.245126}

\bibitem{Mendona2013}
Mendon\c{c}a J~R~G 2013 {\em Journal of Physics A: Mathematical and
  Theoretical\/} {\bf 46} 295001 ISSN 1751-8121
  \urlprefix\url{http://dx.doi.org/10.1088/1751-8113/46/29/295001}

\bibitem{costa2025emergence}
Costa J, Ribeiro P and De~Luca A 2025 {\em arXiv:2504.00188\/}
  \urlprefix\url{https://arxiv.org/pdf/2504.00188}

\bibitem{Li2026}
Li T, Guan S and Suo X 2026 In preparation

\end{thebibliography}

\end{document}